\documentclass[conference]{IEEEtran}
\IEEEoverridecommandlockouts
\usepackage{cite}
\usepackage{amsmath,amssymb,amsfonts}
\usepackage{algorithmic}
\usepackage{graphicx}
\usepackage{textcomp}
\usepackage{xcolor}
\usepackage{balance}
\def\BibTeX{{\rm B\kern-.05em{\sc i\kern-.025em b}\kern-.08em
    T\kern-.1667em\lower.7ex\hbox{E}\kern-.125emX}}
    
\usepackage{url}
\usepackage{enumitem}
\usepackage{subcaption}
\usepackage{pifont}
\newcommand{\cmark}{\ding{51}}%

\usepackage{booktabs,tabularx}
\newcolumntype{x}{>{\raggedright\arraybackslash}X}
\include{balance}

\begin{document}

\title{Captcha Attack: Turning Captchas Against Humanity}


\author{\IEEEauthorblockN{1\textsuperscript{st} Mauro Conti}
\IEEEauthorblockA{\textit{Department of Mathematics} \\
\textit{name of organization (of Aff.)}\\
Padua, Italy \\
conti@math.unipd.it}
\and
\IEEEauthorblockN{2\textsuperscript{nd} Luca Pajola}
\IEEEauthorblockA{\textit{Department of Mathematics} \\
\textit{University of Padua}\\
Padua, Italy \\
pajola@math.unipd.it}
\and
\IEEEauthorblockN{3\textsuperscript{rd} Pier Paolo Tricomi}
\IEEEauthorblockA{\textit{Department of Mathematics} \\
\textit{University of Padua}\\
Padua, Italy \\
tricomi@math.unipd.it}}

\maketitle

\begin{abstract}
Nowadays, people generate and share massive content on online platforms (e.g., social networks, blogs).
In 2021, the 1.9 billion daily active Facebook users posted around 150 thousand photos every minute. 
Content moderators constantly monitor these online platforms to prevent the spreading of inappropriate content (e.g., hate speech, nudity images). Based on deep learning (DL) advances, Automatic Content Moderators (ACM) help human moderators handle high data volume. 
Despite their advantages, attackers can exploit weaknesses of DL components (e.g., preprocessing, model) to affect their performance. 
Therefore, an attacker can leverage such techniques to spread inappropriate content by evading ACM.
\par
While surfing the Web, we observed users posting hateful or sexually explicit content with simple `obfuscation' techniques that recall classical textual captchas. Through our investigation, we revisit the concept of textual captchas and propose a taxonomy of obfuscations applied by Web users.
Then, inspired by these observations, we propose CAPA (CAPtcha Attack), an adversarial technique that allows users to spread inappropriate text online by evading ACM controls.
CAPA, by generating custom textual CAPTCHAs, exploits ACM's careless design implementations and internal procedures vulnerabilities.
We test our attack on real-world ACM, and the results confirm the ferocity of our simple yet effective attack, reaching up to a 100\% evasion success in most cases.
At the same time, we demonstrate the difficulties in designing CAPA mitigations, opening new challenges in CAPTCHAs research area. 
As a result of our experiments, we found that outlier detection approaches were effective in spotting captchas over usual social media content, exceeding 80\% of the F1-score.

\end{abstract}


\section{Introduction}\label{sec.introduction}

Over the last thirty years, the shape and purpose of the Web have changed drastically. If people were initially limited to view content passively, interactions over the Internet are dominant nowadays. The so-called Web 2.0, or Social Web, emphasizes user-generated content and stimulates a participatory culture, giving birth to virtual communities. The Social Web comes in various forms, such as Wikis, Blogs, Social Sharing Platforms (e.g., YouTube), and Online Social Networks (OSN). Users can exchange information, build relationships, and communicate with each other. Even e-commerce websites are populated by user-generated content, such as product reviews and photos. 
\par
To preserve and grant users a safe environment, human content moderators operate controlling users' activity and removing malicious content, such as toxic and hateful messages or violent and sexually explicit images or videos. 
The more content users generate, the more moderators are needed. This applies especially in OSN, in which most of the content is the one published by their users. OSN like Facebook, Instagram, and Twitter are online platforms where users connect, share ideas, opinions, and personal life events. There are more than 4.2 Billion active social media users~\cite{socialStats}, with Facebook leading with around 1.9 Billion daily active users~\cite{FBstatista}.
Moreover, users interact at an incredible pace through posts' reactions (e.g., like, love, do not like), comments, and sharing. For example, Instagram and Facebook count about 350 thousand stories and 150 thousand photos posted, respectively, every minute~\cite{socialStats}. 
Human moderators cannot keep up with such a rate, and the need for automated moderators is then increasing. Besides, content such as child pornography, hate-filled messages, and gratuitous violence can cause considerable psychological risks to the human moderators~\cite{arsht2018human}.
\par
\begin{figure}
    \centering
    \includegraphics[width = .6\linewidth]{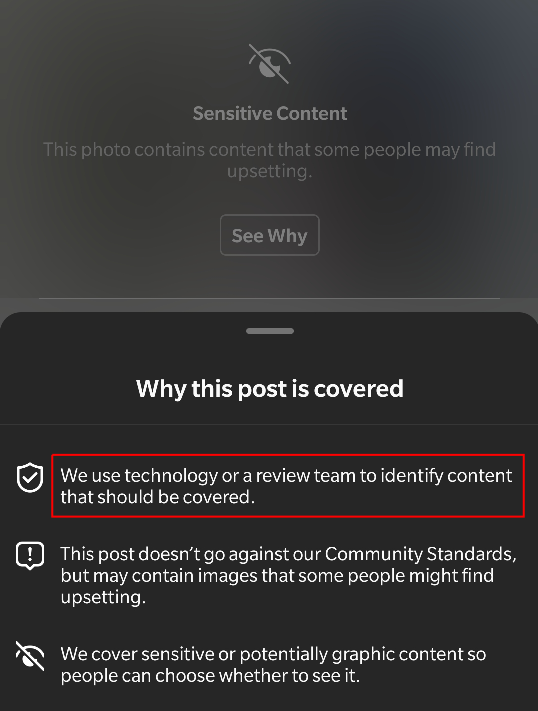}
    \caption{Instagram alert of sensitive content.}
    \label{fig:ig_content}
\end{figure}

Researchers and companies started developing automatic tools to face the massive amount of user-generated content and tackle the problem of ``malicious content detection''. 
These tools are mainly solved with data-driven approaches, e.g., machine learning (ML). Some examples are the hate-speech detectors~\cite{warner2012detecting,djuric2015hate}, which aim to detect sentences targeting a group of people based on their characteristics or race. 
Thus, OSN like Facebook adopt Automatic Content Moderators (ACM) to help human moderators during their monitoring tasks. In particular, as reported by \textit{TheVerge}~\cite{fb_ai_the_verge}, Facebook employs ML tools to monitor users' posts to spot potential inappropriate content that human operators will manually review. Such content is either removed or labeled as ``sensitive'', which means users have to explicitly accept to view it. 
Instagram recently adopted a similar system~\cite{ig_content}, stating that technology and humans are used to identify sensitive content (Figure~\ref{fig:ig_content}).
\par
\begin{figure}[!ht]
    \centering
    \includegraphics[width = 0.7\linewidth]{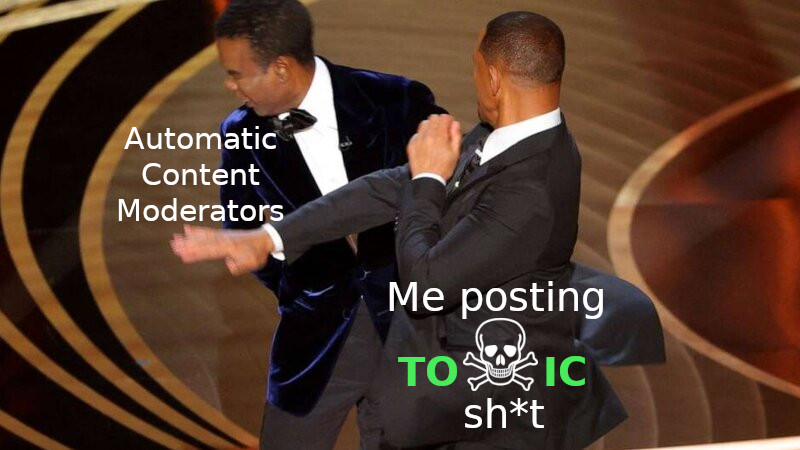}
    \caption{Example of meme with different obfuscations (e.g., typos, letters-shaped objects, hard background).}
    \label{fig:meme-example}
\end{figure}

At the same time, the security of such tools is fundamental to avoid malicious users spreading unauthorized content. 
For example, if ACM does not detect inappropriate content, human moderators will not control the content, and it will spread on the platform. 
Thus, only users reports could alarm human moderators, but, in this way, the content might have already harmed those that saw it.
ML-based solutions are vulnerable to \textit{evasion attacks}, where the attacker feeds models with crafted samples aiming to affect models' predictions~\cite{biggio2013evasion}. 
However, state-of-the-art attacks usually suffer limitations such as requiring deep knowledge of the attacked system, are not easy to implement, and focus ML algorithms' weaknesses rather than complete applications vulnerabilities. 
Indeed, ML applications also contain preprocessing functions that, if exploited, can affect ML models' decisions~\cite{xiao2019seeing,pajola2021fall}.
\paragraph{Contribution}
Surfing online platforms like Instagram and Facebook, we noticed peculiar harmful posts, obfuscated with techniques that allow human understanding while evading ACM.
The observed techniques recall those used to generate textual captchas\footnote{We use the lowercase form of CAPTCHA to improve paper's readability.} (e.g., occluding items, rotations). Figure~\ref{fig:meme-example} shows an example of a meme containing some of these obfuscations (e.g., typos, letters-shaped objects, hard background).
Thus, we investigated this phenomenon and harvested such adversarial samples from the wild to define a taxonomy of obfuscation techniques. 
This contribution seeks to revisit the classical concept of textual captchas (defences used in web platforms to distinguish humans from machines), and define a new type of captcha generated by humans to evade machines.
\par
\begin{figure}[!htbp]
    \centering
    \includegraphics[width = .6\linewidth]{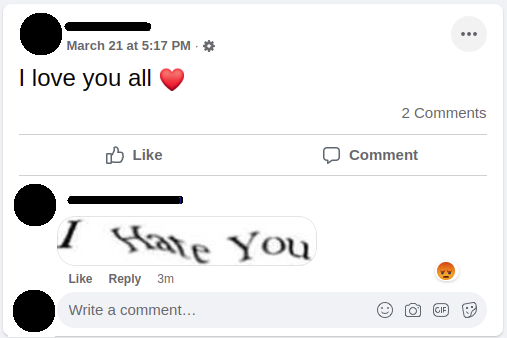}
    \caption{Example of a custom textual captcha containing the phrase ``I Hate You''.}
    \label{fig:ex_captcha}
\end{figure}

Inspired by the categories of our taxonomy, 
we propose a new attack aiming to spread adversarial sentences without being detected by ACM. 
Examples of such sentences are inappropriate words (e.g., offenses, hate speech).
The proposed attack consists of writing words with textual captcha-based strategies, which, per definition, are challenges that only human beings can solve.
We named our attack CAPA\footnote{The name of our attack is a quote to Caparezza, an Italian singer famous for his lyrics rich with puns.} (i.e., CAPtcha Attack).
While textual captchas usually contain random characters, in this work, we investigate what might happen if an attacker can instead choose the characters (i.e., words) that compose a textual captcha.  
An example is reported in Figure~\ref{fig:ex_captcha}, which is generated with the free online tool \textit{Fake Captcha}\footnote{\url{https://fakecaptcha.com/}}.
An attacker can thus spread custom textual captchas on online platforms, threatening ACM detection performance. 

\par
While generating CAPA is simple and effective, we show that defining a robust defense mechanism is anything but trivial. In our experiments, we discuss potential failure of supervised approaches and proposed an unsupervised defense based on outliers detection, reaching the 80\% F1-score. 
We summarize our contribution as follows:
\begin{enumerate}[noitemsep,topsep=0pt,parsep=0pt,partopsep=0pt]
    \item We define a taxonomy of modern obfuscation techniques used by web users to evade ACM;
    \item We propose an evasion attack called  CAPA,  a novel threat that allows users to spread dangerous textual content which eludes ACM;
    \item We demonstrate that CAPA (i.e., custom textual captchas) have high evasion capability on real-world ACM\footnote{We test ACM APIs to avoid the spread of inappropriate content.}; 
    \item We discuss challenges of possible defense mechanism to prevent CAPA. We further propose an outlier detection-based defense that reaches up to 80\% of F1-score.
\end{enumerate}

\paragraph{Paper organization}
The paper is organized as follows. In Section~\ref{sec.rel_works}, we briefly describe the concepts required to understand the rest of the paper thoroughly. 
Section~\ref{sec:taxonomy} defines a taxonomy of obfuscation techniques harvested from the wild.
In Section~\ref{sec.attack_model}, we formalize CAPA attack.
In Section~\ref{sec.dataset} we describe the dataset generated with CAPA, followed by Section~\ref{sec.attack}, where we show results of the attack in real-world automatic content moderators. 
We then propose a possible countermeasure in Section~\ref{sec.defense}.
We finally conclude our paper and propose future directions in Section~\ref{sec.Conclusions}.
\section{Background \& Related Works}\label{sec.rel_works}
This section presents theoretical concepts with related works required to understand the rest of the paper entirely. In this section, we discuss about ML-based applications security (Section~\ref{sub.rw_ml}), the research area of Automatic Content Moderators (Section~\ref{sub.rw_cm}), and finally the history of captchas (Section~\ref{sub.rw_cap}). 
\subsection{Security of Machine Learning Applications}\label{sub.rw_ml}
%
%

\par
ML applications like automatic content moderators need to deal with real-world challenges, offering at the same time high performance and attack resiliency. 
Therefore, when considering the application security, we need to consider all of the components of such pipelines, like preprocessing function, machine learning algorithms, and developing libraries (e.g., PyTorch, Scikit-learn). 
In general, an adversary's goal is to control and affect ML application decisions through the definition of \textit{adversarial samples}.
\par
In the literature, to the best of our knowledge, the focus has been on identifying threats related to the ML algorithms rather than the entire pipeline. 
In the latter, an attacker leverages the pipeline's weaknesses (e.g., application flow, libraries bugs) to achieve his/her goal (e.g., evasion).
For example, in~\cite{8424643}, the authors exploit software implementation bugs of popular deep learning frameworks (e.g., Caffe, TensorFlow) to launch denial-of-service (DoS) attacks that crash the applications. 
In the image domain, the \textit{camouflage attack} exploits image scaling logic to alter the semantic meaning of the image after the transformation~\cite{xiao2019seeing}. 
In the text domain, \textit{ZeW} attack affects text representations by injecting non-printable UNICODE characters~\cite{pajola2021fall}. 
On the opposite, vulnerabilities related to ML algorithms are widely explored. We find different classes of attacks, such as the \textit{evasion attack}, where the attacker defines malicious samples that fools a target classifier~\cite{biggio2013evasion,goodfellow2015explaining}, and the \textit{poisoning attack}, where the attacker affects model performance if he/she has access to the training data~\cite{barreno2006can,rubinstein2009antidote}.
\par
Finally, in this work, we focus on adversarial sentences spreading through images.
Given captchas deceiving nature, we can categorize our attack as a \textit{cross-modal} attack on \textit{Optical Character Recognition} (OCR)~\cite{10.1145/3374217}. 
OCR are tools aiming to extract text from images. 
Baseline adversarial attacks on OCR use different strategies like noise and watermark addition~\cite{fawa,chen2020attacking}. 
These attacks are optimized to fool a target model. 
In contrast, our proposed attack leverages captchas that are a natural antagonist of OCR by definition. 
\par
Thus, the proposed attack CAPA is not optimized to fool ACM machine learning algorithms but rather to affect earlier stages, such as the text extraction from images using OCR.
\subsection{Content Moderators}\label{sub.rw_cm}
Online platforms use human moderators to monitor content shared in their virtual environment. 
Their role is fundamental to block any malicious content before spreading.  
However, their efficiency is limited by the many users and interactions that a platform presents daily. 
To overcome this issue, companies started developing automatic tools. 
As stated in~\cite{doi:10.1177/2053951720943234}, ``\textit{the major platforms dream of software that can identify hate speech, porn, or threats more quickly and
more fairly than human reviewers, before the offending content is ever seen.}''
For example, Facebook uses both human and automatic content moderators: ML filters flag potentially harmful content, and clear-cut cases are removed automatically, while the rest are processed by human operators~\cite{fb_ai_the_verge}.
\par
Human and automatic content moderators need to deal with multimodal content such as text, image, video, and audio. 
We can thus find several moderator tools based on the aim and source type. 
A popular and widely studied application is \textit{hate speech detection}. 
While these tools mainly focus on textual contents with NLP-based solutions~\cite{schmidt-wiegand-2017-survey}, only recently the attention moved on the multimodal representations (e.g., text inside images). 
For example, a new popular trend is the hateful meme detection~\cite{Kiela2020TheHM,Gomez_2020_WACV,velioglu2020detecting}, where the ACM combines images and textual information to address the task.
Finally, online platforms are often visited not only by adults but by children as well. Image and video can contain contents that are not appropriate for such a young audience. Examples are \textit{violent} and \textit{sexually explicit} content detectors~\cite{SUN201943,10.1145/3098954.3104051}.
\par
Generally, users benefit from automatic content moderators since they allow an improvement of platforms' quality.  
Nevertheless, popular platforms are populated by malicious users who aim to disrupt such ecosystems. For example, in 2016, a group of users affects \textit{Tay}'s response behavior, a Microsoft chatbot; this tool was shut down after it started spreading hateful tweets~\cite{10.1145/3144592.3144598}. 
At the same time, automatic content moderators have been proved to be vulnerable to adversarial attacks. In~\cite{yuan2019stealthy}, the authors highlight the attention on \textit{real-world} adversarial techniques on sexually explicit detectors. Here, cyber-criminals used simple image transformations (e.g., rotation, noise addition) to spread porn images on online platforms without being detected. Similarly, in~\cite{10.1145/3270101.3270103}, the authors presented ``all you need is love'', showing that the popular toxic comments detector \textit{Google Perspective}\footnote{\url{www.perspectiveapi.com}} could be affected by the addition of simple typos and love words. 
\subsection{CAPTCHA}\label{sub.rw_cap}
A CAPTCHA (Completely Automated Public Turning Test to tell Computers and
Humans Apart) is a test to distinguish between humans and computers (e.g., bots, automated users). First examples appear in 2000, designed by Von Ahn et al.~\cite{von2003captcha}, to check whether web requests were coming from humans, improving the security of websites, such as by preventing spam, protecting users' registration, and limiting email address scraping. The first generation of captchas was based on text, altered by rotations, distortions, or wavings, to be hardly readable by a machine (e.g., OCR) but simple for humans.
With the advancements in AI technology, text-based captchas began to be solved, with a significant decline in 2014, when Google demonstrated that even the most complicated variants could be easily broken~\cite{goodfellow2013multi}. The security weaknesses related to text-based captchas led the research community to develop new techniques. Image-based ones were the first alternatives, followed by audio, video, and puzzle-based captchas~\cite{singh2014survey}. In general, their evolution follows the advancements of technology to break them~\cite{guerar2021gotta}. Even if the text-based captchas security has been proved to be inefficient, they are still preferred by many users because of familiarity and sense of security and control~\cite{krol2016better}.
\par
The research community put much effort into solving (or breaking) text-based captchas (the type used in our attack). Their robustness has been shown to heavily rely on the difficulty of finding where the character is, i.e., \textit{segmentation}, rather than what character it is, i.e., \textit{recognition}~\cite{chellapilla2005computers}. The breaking methods evolved from algorithmic techniques~\cite{mori2003recognizing,yan2008low} to machine learning based approaches~\cite{bursztein2014end,yan2016simple,goodfellow2013multi}.
\par
To the best of our knowledge, there are no prior works in the literature to detect whether an image is a textual captcha. 
A possible explanation is that in attacking a website or a web service, the attacker usually knows the phase when a captcha is required and its schema, and for this reason, the research community focused on the breaking path rather than their recognition. 
Recognizing if an image contain textual captchas could be an effective CAPA defense. 
Thus, we pose a new problem of distinguishing a textual captcha from other real-world objects in images. 
\section{The new era of textual captcha}\label{sec:taxonomy}
In this section we investigate obfuscation techniques we found on OSN. An overview is given in Section~\ref{sub.tax-over}. 
We then discuss the three macro-level of obfuscations: OCR-failures (Section~\ref{sub.tax-ocrf} and classifier-failures (Section~\ref{sub.tax-cf}).

\subsection{Challenges from OSN's users: a Taxonomy}\label{sub.tax-over}
\textit{What is a captcha?} CAPTCHA acronym is self-explanatory: ``Completely Automated Public Turing test to tell Computers and Humans Apart''. In other words, a captcha is a task that a human can accomplish, but the machine not. 
Our perception of captchas is highly biased by their adoption as a defense mechanism in web services. For example, when we think of a textual captcha, we imagine an artificial image with random characters that are transformed with several effects (e.g., blur). 
\par
On the opposite, the evolution of textual captchas can be easily found on the web, surfing social networks like Facebook and Instagram.
While looking at memes published in popular accounts, we noticed that those containing potential textual harmful contents (e.g., swear words, hate speech, sexually-related) had been somehow \textit{obfuscated} by their owners. 
With the term obfuscation, we mean a set of transformations that potentially mine ACM actions. 
The obfuscation we found are somehow an evolution of traditional textual captchas styles, with an offensive use of the technology.
\par

\begin{figure*}[!ht]
    \centering
    \includegraphics[width=.95\textwidth]{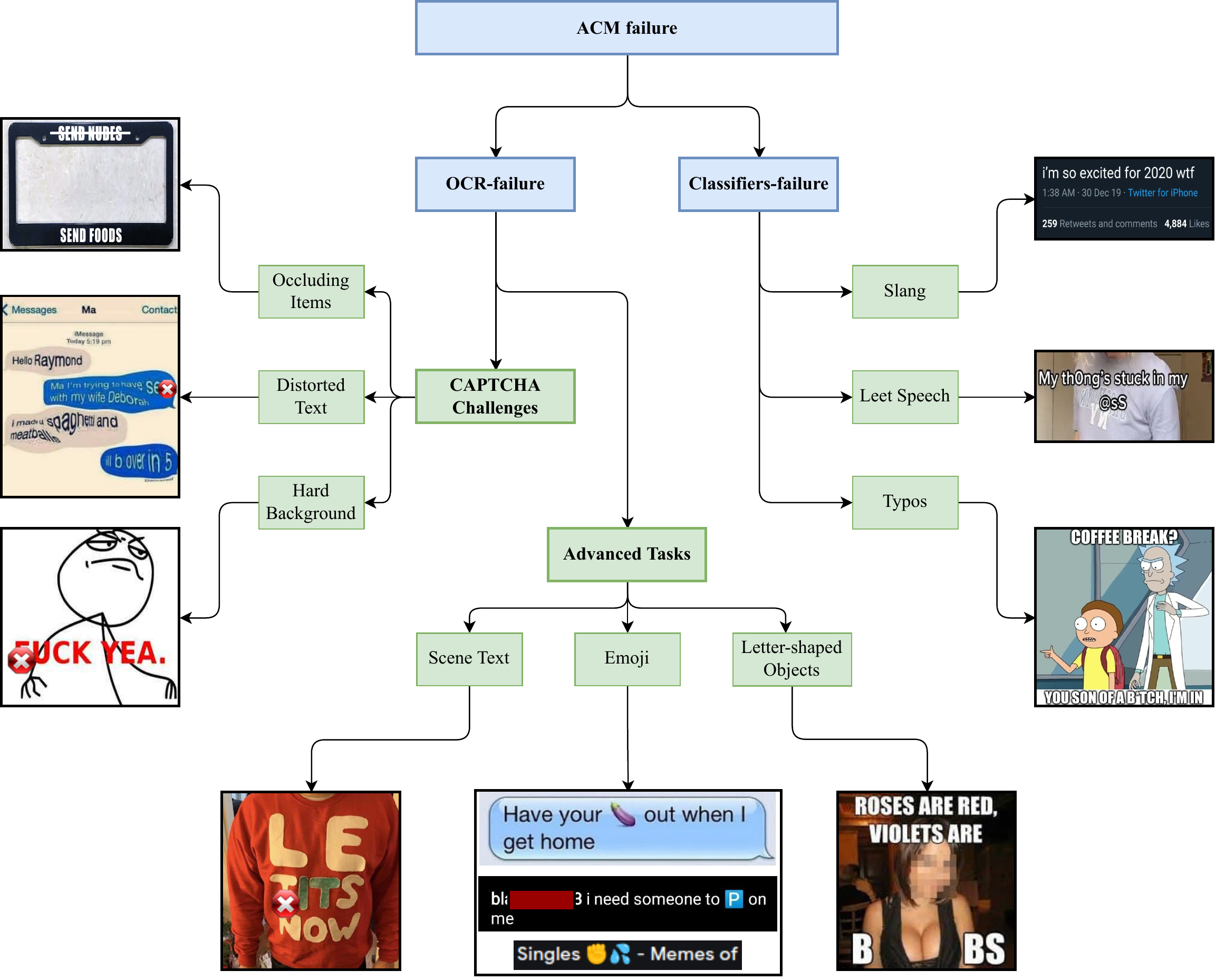}
    \caption{Obfuscation techniques we identified in online social networks. Blue boxes represents the ACM component that might fails. Green boxes represent different obfuscation techniques. We censored explicit or harmful contents with the red symbol.}
    \label{fig:taxonomy}
\end{figure*}

We thus decided to investigate the nature of such posts profoundly. Since the language can be a barrier to identify elements such as typos, slang, or double meanings, we focused only on posts in English or Italian, which we could fully understand. In this stage, we targeted Instagram, one of teenagers' most used platforms. We selected three well-known English pages, three famous Italian pages, and four popular hashtags, all related to memes or adult (potentially harmful) content. We limited our manual inspection to the latest 100 posts for each page since these obfuscation techniques seem to have been adopted only recently.
Given that hashtags convey content from many pages and users, we focused on the latest 1000 English posts without incurring the risk of analyzing old content. Every time we encountered a new potential obfuscation technique, we started keeping track of the posts involving it. Then, we grouped similar obfuscation techniques into categories, producing the final taxonomy presented in Figure~\ref{fig:taxonomy}.
We want to underline that, in this phase, we considered not only potential harmful posts, but any post published in such pages that might mine the correct workflow of ACM.
\par
In the taxonomy, we show samples that are as harmless as possible while maintaining a clear understanding of the obfuscation techniques.
As we will explain later in Section~\ref{sec.attack_model}, an ACM that deals with images containing text (like memes), presents an OCR that extracts the text and then use a set of ML classifiers to analyze the content. 
Thus, the first division of our taxonomy relates to which ACM component might be deceived by the obfuscation: \textit{OCR} or \textit{ML classifiers}. 
In the rest of this section, we explore the various identified cases in detail.
\par
Table~\ref{tab.tax-stat} shows the statistics about the distributions of obfuscation techniques we encountered when creating the taxonomy. We can first notice that `hard background' is present in most of the sources, reaching the 77\% in one case. In general, this seems a trend of new posts, where the text is written on top of a complex background (e.g., real life scene).
Moreover, we noticed that some techniques (i.e., emoji, leet speech, typos, and occluding items) were mainly used to cover sexually explicit content or swear-words.

\begin{table*}[ht!]
\centering
\footnotesize
\def\arraystretch{1}
\caption{Percentage of obfuscation techniques observed in different Instagram sources.}
 \label{tab.tax-stat}
  \begin{tabular}{lccc| ccc| ccc}
    \toprule
        
         &   \multicolumn{3}{c}{\textbf{\textit{CAPTCHA Challenges}}} &
        \multicolumn{3}{c}{\textbf{\textit{Advanced Tasks}}} &
        \multicolumn{3}{c}{\textbf{\textit{Classifier-Failures}}} 
        \\
             
      \cmidrule(lr){2-4} \cmidrule(lr){5-7}  \cmidrule(lr){8-10}
      
    \multicolumn{1}{l}{\textbf{\textit{Source}}}  
    & Occ. Items & Dist. Text & Hard Back.
    & Scene Text & Emoji & LSO
    & Slang & Leet Speech & Typos
    \\
    \hline
    epicfunnypage &	3.0 &	10.0 &	14.0 &	9.0 &	5.0 &	0.0 &	26.0 &	1.0 &	4.0 \\ 
    6.memes.9 &	6.0 &	7.0 &	33.0 &	4.0 &	5.0	& 0.0 & 5.0	& 0.0 &	5.0 \\
    9Gag &	0.0 &	0.0 &	20.0 & 1.0 & 0.0 & 0.0 &	7.0 & 0.0 & 0.0 \\
    partitodisagiato &	11.0 &	0.0 &	77.0 &	1.0 &	23.0 &	0.0 &	3.0 &	15.0 &	4.0 \\
    pastorizianeverdiesreal & 0.0 &	0.0 &	29.0 &	2.0 &	1.0 &	0.0 &	1.0 &	0.0 &	0.0 \\ 
    alpha\_man\_real &	1.0 &	1.0 &	47.0 & 1.0 & 0.0 &0.0 & 1.0	& 0.0 &3.0 \\
    \#naughtymemes &	5.3 &	4.1 &	23.8 &	5.2 &	12.1	& 0.2 &9.7 &3.0 & 5.4 \\
    \#sexualmemes	&	3.2 &	6.1 &	23.2	& 4.7	& 7.3	 & 0.0 &	15.2 	& 0.4	& 5.5 \\
    \#nsfwmemes	&	7.0	& 7.1	& 31.8 & 0.8 	& 4.2 &	0.0 &	14.5	& 2.1 &	5.9 \\
    \#adultmemes &	1.3 &	3.2 &	18.4 &	5.9 &	4.7	& 0.4 &	10.0 &	4.3 &	5.3 \\
    \bottomrule
\end{tabular}
\end{table*}

\subsection{OCR-failure}\label{sub.tax-ocrf}
OCR-level obfuscations aim to disrupt or affect the text extraction phase from images. 
We identified two sub-family of techniques: \textit{advanced task} for OCR and \textit{CAPTCHA challenges}. 

\subsubsection{Advanced Tasks for OCR}
With \textit{advanced tasks} we mean a set of applications that differ from the classic document extraction and pose more challenges for OCR. 
For example, \textit{scene text} recognition is an area that gained popularity in the last few years~\cite{long2021scene}.
This task consists of detecting and extracting text from real-life scenes (e.g., a road sign, T-shirt). 
\par
Another exciting challenge is letter-shaped objects, i.e., images whose shape recall a specific alphabet letter. OCR might not recognize the correct character, resulting in an erroneous extraction. 
This task is not yet discussed in OCR literature to the best of our knowledge. 
\par
We conclude with the family of \textit{emoji} obfuscations. In Figure~\ref{fig:taxonomy}, we show three typical examples of emoji obfuscations. On the top, the text contains an eggplant with a visual double-meaning (i.e., referring to a penis). 
In the middle, the P-emoji is used with a phonetic deception (i.e., P can be read as `pee').
On the bottom, two emoji are combined to represent a sexual action. 


\subsubsection{CAPTCHA Challenges}
Textual CAPTCHA challenges represent obfuscations usually adopted by textual captchas. 
Such transformations are \textit{hard background}, \textit{distorted text}, and \textit{occluding items}.
While we classified these three obfuscations as a stand-alone, they are usually blended with other obfuscations we presented in the taxonomy. 

\subsection{Classifier-failures}\label{sub.tax-cf}
ML-level obfuscations contains techniques that, while allowing a proper textual extraction, mine the correct functioning of ML classifiers. 
These techniques are similar to those presented in~\cite{10.1145/3270101.3270103}: slang, leet speech, and typos. 
The first category relates to post that contains slang terms (e.g., wtf $\rightarrow$ what the f*ck). 
The second class is the \textit{leet speech}, where some characters are replaced with other visually similar ones, e.g., $a*s \rightarrow @sS$.
The last class relates to text with \textit{typos} or grammatical mistakes, i.e., images containing misspelled words that, however, can be comprehended by human readers. 
In the example, we show a meme that contains a sentence with a swear word where the letter `i' is replaced by `*'.
\section{Our attack: CAPA}\label{sec.attack_model}
In this section, we present the proposed attack by first analyzing the target system (Section~\ref{sub.am_as}). Then, we discuss the attack motivation (Section~\ref{subsec:motivation}), and formalize its execution (Section~\ref{sub.am_pa}).

%
%
%
\subsection{Target System}\label{sub.am_as}
In this work, we aim to highlight potential ACM vulnerabilities with the final goal of making them more secure. 
In particular, we focus on the (easy) generation of content that could potentially evade any ACM activity. 
We build our attack by considering that, in online platforms, users mainly communicate with textual and image interactions. 
For example, on Facebook, users can use textual and image comments under others' posts.
\par
As introduced in Section~\ref{sub.rw_cm}, only a few works focus on identifying toxic contents based on the combination of both textual (extractable with OCR) and image information~\cite{Kiela2020TheHM,Gomez_2020_WACV,velioglu2020detecting}. 
Based on such state-of-the-art, we can abstract the execution process of ACM that deal with both textual and image content. 

\begin{figure*}[!htbp]
    \centering
    \includegraphics[width = 0.6\linewidth]{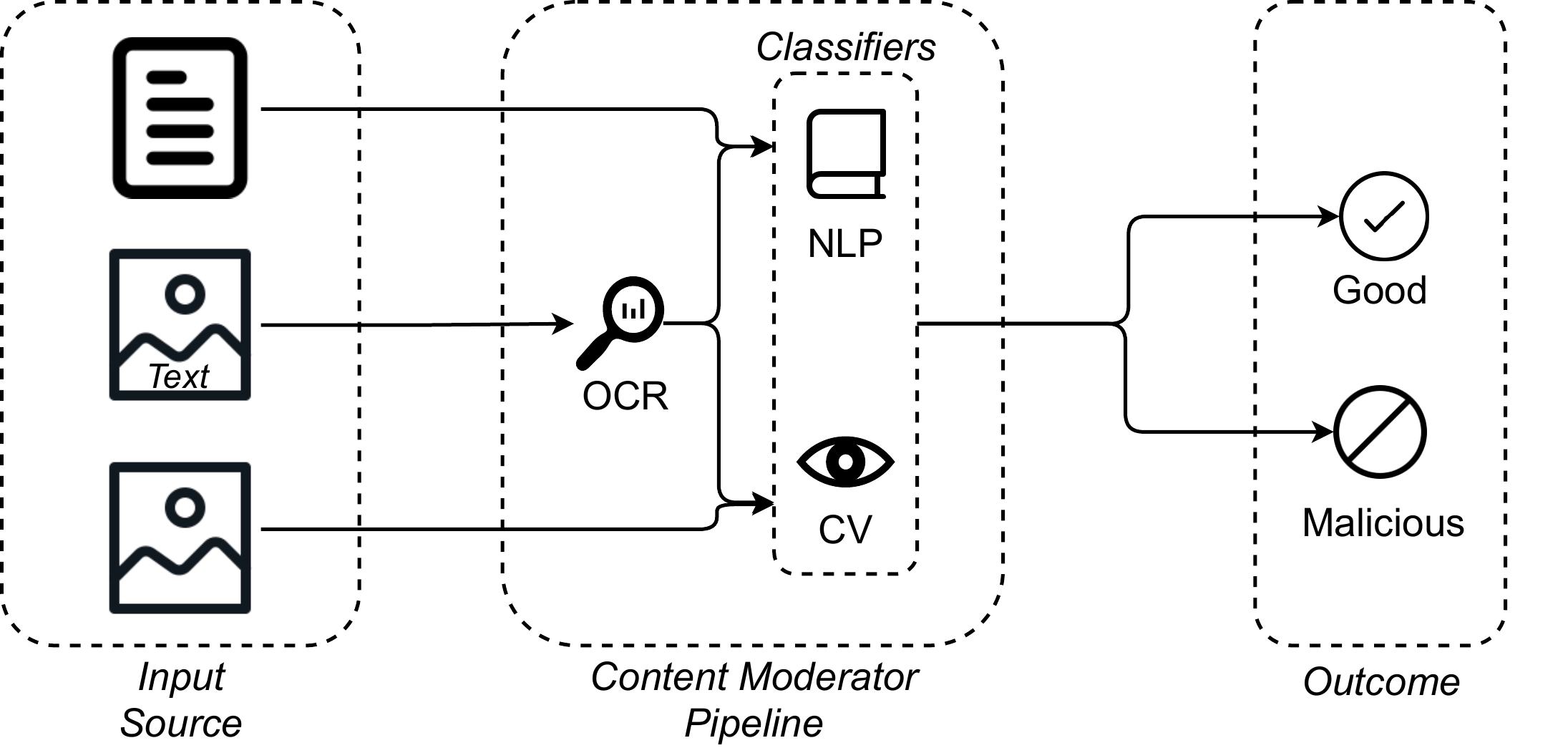}
    \caption{Overview of a content moderator in the text and image domains.}
    \label{fig:cross_domain_moderator}
\end{figure*}

Figure~\ref{fig:cross_domain_moderator} shows such a pipeline. 
In particular, we can find four possible inputs, as following described.
\begin{itemize}[noitemsep,topsep=0pt,parsep=0pt,partopsep=0pt]
    \item \textit{Textual contents} (e.g., comments). The ACM analyzes the content using Natural Language Processing (NLP) tools. 
    \item \textit{Image content} (e.g., photos). The ACM analyzes the content using Computer Vision (CV) tools.  
    \item \textit{Text with image content} (e.g., posts). The ACM analyzes the content with a set of cooperative NLP and CV tools.
    \item \textit{Images containing text content} (e.g., memes). Like the previous case, the ACM analyzes the content with a set of NLP and CV tools in cooperation. Moreover, this case requires additional preprocessing steps to extract textual information from images (e.g., OCR). 
\end{itemize}
How NLP and CV models should cooperate in multimodal cases is out of the scope of this paper. 
In contrast, we highlight that when images contain text (i.e., the case of textual captchas), a vital pipeline phase is left to OCR to extract information to feed NLP models.
\par
Note that the proposed pipeline only presents a general overview of how an ideal ACM should work.
While different companies can adopt and develop different ACM, our proposed pipeline can still faithfully describe their workflows since we do not discuss how to implement specific operations.  

\subsection{Attack Motivation}
\label{subsec:motivation}
In Section~\ref{sec:taxonomy}, we presented examples of real-life obfuscations we spotted on social networks like Facebook, Instagram, and Twitter. 
Among these posts, we saw several extremely inappropriate ones (e.g., sexually explicit, hateful sentences) obfuscated with one or more techniques.
Studying ACM' behavior in the presence of such `adversarial' samples would highlight ACM' weaknesses. 
Behind these obfuscations, we always find the same rationale: people are trying to create content that can be easily understood by humans but is challenging for machines. 
In other words, people are forging a new generation of textual captchas, which are, per definition, the natural antagonist of OCR.
This is the first attempt to use textual captchas as a malicious vehicle to the best of our knowledge.
\par
An ideal way to study how ACM would behave with these malicious samples would require collecting a vast number of them. 
However, we find three major challenges to collect such dataset: (1)
these obfuscations seem novel and a direct consequence of the recently adoption of ACM in OSNs~\cite{fb_ai_the_verge}, resulting in a limited number of samples; (2) there are many variants or ways to produce an obfuscation, making the problem of limiting samples worse; (3) an automatic tool to detect such posts currently does not exist. 
\par
To address the previously listed issues and to effectively evaluate current real-world ACM' robustness, we focus on a the automatic generation of classic textual captchas containing custom words (e.g., see Figure~\ref{fig:ex_captcha}). We call this attack CAPA: CAPtcha Attack.
Custom textual captchas can be considered a broad sub-category of obfuscation techniques proposed in our taxonomy (Section~\ref{sec:taxonomy}), with the following benefits:
\begin{itemize}[noitemsep,topsep=0pt,parsep=0pt,partopsep=0pt]
    \item Given a set of captcha styles, we can generate an arbitrary number of samples;
    \item The generated samples represent a simplified version of real-life posts since they do not contain any visual aspect that might affect CV classifiers;
    \item Classic textual captchas have been widely investigated in the literature, and thus the knowledge acquired so far might help counter CAPA and, more in general, the obfuscation techniques we discussed in our taxonomy (Section~\ref{sec:taxonomy});
    \item They can be the basis for an attack that potentially disrupt any OCR.
\end{itemize}


\subsection{Attack Execution}\label{sub.am_pa}
In this section, we briefly describe the process of generating a textual captcha, which is the basis of CAPA execution. 
In particular, given an harmful custom textual sample $x$, and an automatic content moderator $M$, we aim to identify a transformation function $T$ such that:
\begin{equation}
\begin{split}
    &M(x) = c_i,\\
    &M(T(x)) = c_j, 
\end{split}
\end{equation}
where $c_i$ is the offensive class, and $c_j$ the non offensive one. 
The function $T$ should satisfy the following properties.
\begin{enumerate}[noitemsep,topsep=0pt,parsep=0pt,partopsep=0pt]
    \item \textit{Easy to deploy}. This would open to a broad target of possible adversaries, not only people highly skilled in computer science.
    \item \textit{Target model agnostic}. The transformation should be independent of the target system, i.e., the process $T$ is not mathematically optimized to fool a specific ACM $M$, but rather any ACM. This would make the attack stronger and effective to different unknown ACM. 
    \item \textit{Effective}. The attack should be successful with high confidence. This is desirable since online platforms follow strict policies for inappropriate content sharing, e.g., suspension or account ban. 
\end{enumerate}
\par
The seek of a transformation function $T$ that satisfies the described properties starts with the following insights: 
\begin{enumerate}[noitemsep,topsep=0pt,parsep=0pt,partopsep=0pt]
    \item CV-based applications can be easily fooled, as shown with simple yet effective image manipulations~\cite{yuan2019stealthy};
    \item adversarial perturbations in the text-domain present a trade-off between readability and attack success rate~\cite{10.1145/3374217}.
\end{enumerate} 
We thus decided to combine the text with the image domain, resulting in images displaying text.
Therefore, malicious users aiming to spread undetectable comments can post images containing the desired words. 
This domain-change transformation $T_1$ represents our first deceptive layer.
\par
While analyses on the text and image contents follow standard predictions, the case of text contained in images might represent a gray area since it involves additional operations such as text extraction and the cooperation between NLP and CV machine learning algorithms.  
If an online platform does not explicitly develop an ACM handling such cases, there is a high chance that malicious content $T_1(x)$ can evade detection mechanisms. 
We explore this scenario in Section~\ref{sub.ar_c1}.
\par
If we consider proper implementations of automatic content moderators (see  Figure~\ref{fig:cross_domain_moderator}), setting $T = T_1$ might not be sufficient to guarantee complete attack effectiveness. 
Thus, we add typical manipulation and distortion of classic textual captchas, producing images with similar properties of the ones presented in Figure~\ref{fig:taxonomy}.
For example, we noticed most of posts (e.g., memes, Instagram reels) present an hard background. 
A customizable textual captcha can be seen as a function composition:
\begin{equation}\label{eq.attack}
    T = T_n(...(T_2(T_1(x)))),  
\end{equation}
where $T_1$ represents the domain transfer function, while the set $[T_2, ..., T_n]$ is the combination of image transformations to generate the captcha, and $x$ is the given sentence. 
As reported in~\cite{ye2018yet}, popular transformations can be applied at the background (e.g., solid, complex, noisy), character (e.g., font, size, color, rotation, distortion), and word level (e.g., character overlapping, occluding lines, waving, noise).
\par
\begin{figure*}[ht!]
    \centering
    \includegraphics[width = .8\linewidth]{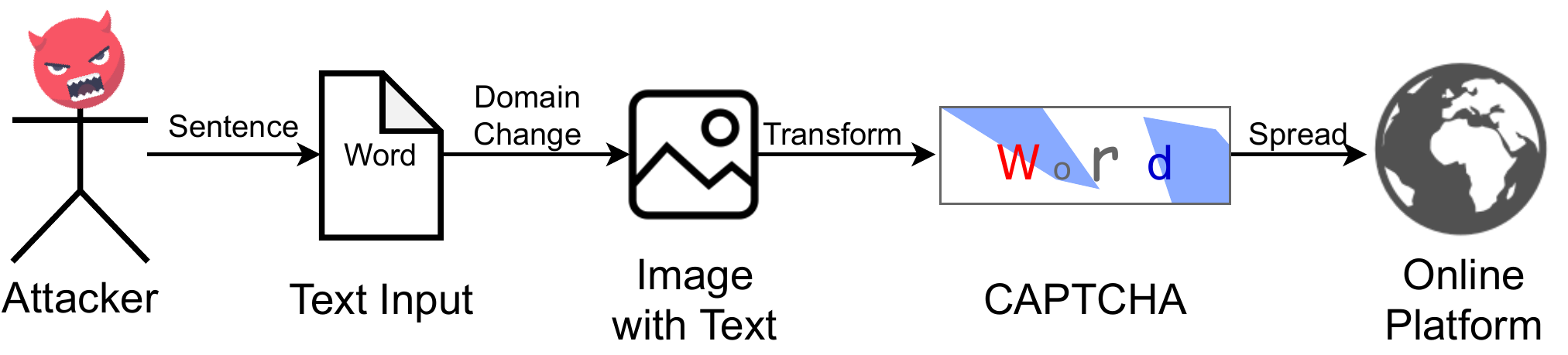}
    \caption{Overview of CAPA execution pipeline.}
    \label{fig:att_pipeline}
\end{figure*}

The generation process we just described is well-known to the state-of-the-art. While this process does not constitute a part of the novelty of this work, in contrast, the usage of captchas from defense solutions to attack vector is, to the best of our knowledge, a novel and unexplored threat. 
Figure~\ref{fig:att_pipeline} shows an overview of the attack execution.
$T$ satisfies the requirements presented at the beginning of this section.
\begin{itemize}
    \item \textit{Easy to deploy}. The generation of textual captchas is trivial. Moreover, malicious users can use several online tools to generate custom textual captchas.
    \item \textit{Target model agnostic}. The captchas generation process does not take into account any information of victims' ACM. 
    \item \textit{Effective}. From a theoretical perspective, the usage of captchas should guarantee a high evasion rate. We demostrate CAPA effectiveness in Section~\ref{sec.attack}.
\end{itemize}

The proposed attack can exploit the following target ACM' weaknesses:
\begin{enumerate}[noitemsep,topsep=0pt,parsep=0pt,partopsep=0pt]
    \item \textit{Unimplemented detection case}. Text inside images is a grey area between CV and NLP. The implementation of cross-domains ACM is not trivial and is not widely explored in literature. ACM not implementing such a scenario will miss images with harmful plain text.
    \item \textit{Text extraction phase}. If ACM deploys the monitoring of multimodal contents, a pipeline key phase is the text extraction. 
    OCR usually handle this operation.
    OCR extraction from textual captchas might result in noisy inputs that feed NLP models and thus affecting their predictions. 
\end{enumerate}


\section{CAPA Dataset}\label{sec.dataset}
In this section, we describe the dataset used to test our proposed attack's performance (Section~\ref{sub:generation}).
Then, we evaluate its readability with a user study (Section~\ref{sec:user-study}).

\subsection{Dataset Generation}\label{sub:generation}
As introduced in Section~\ref{sub.rw_cm}, a popular and essential ACM role is the identification of hateful messages on online platforms. 
Thus, an example of a possible attacker's goal is to let hateful messages being undetected by ACM. 
We build our dataset with potential hateful textual captchas. 
\par
We retrieve a list of frequent English words associated with hateful sentences from \textit{Hatebase.org}~\cite{hatebase}, for a total of 1383 samples. 
From this list, we maintain only those samples that, as stand-alone, should be banned from online platforms using \textit{Microsoft Content Moderator}\footnote{\url{azure.microsoft.com/en-us/services/cognitive-services/content-moderator}} as our ground truth. 
This API identifies the presence of potential undesired textual content through three probability scores: i.e., sexually explicit content, sexually suggestive content, and offensive language content. 
We consider a sample inappropriate if at least one of these three probability scores is greater than 0.5.
The final list contains 502 samples.
\par
\begin{figure*}[!htbp] 
    \centering
  \begin{subfigure}[b]{0.2\linewidth}
    \centering
    \includegraphics[width=.7\linewidth]{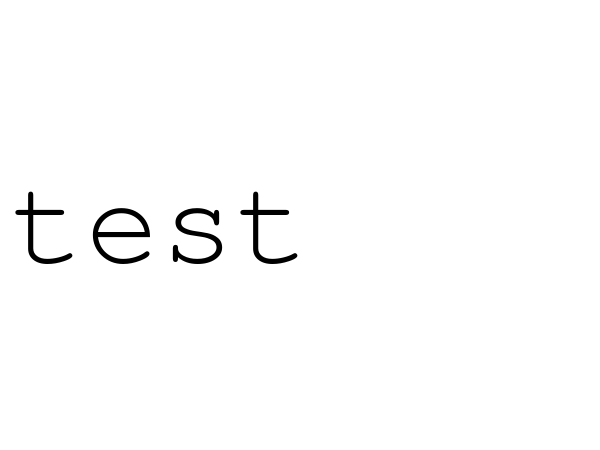} 
    \caption{Clean Image} 
    \label{fig_captchas:a} 
  \end{subfigure}
  \begin{subfigure}[b]{0.2\linewidth}
    \centering
    \includegraphics[width=.7\linewidth]{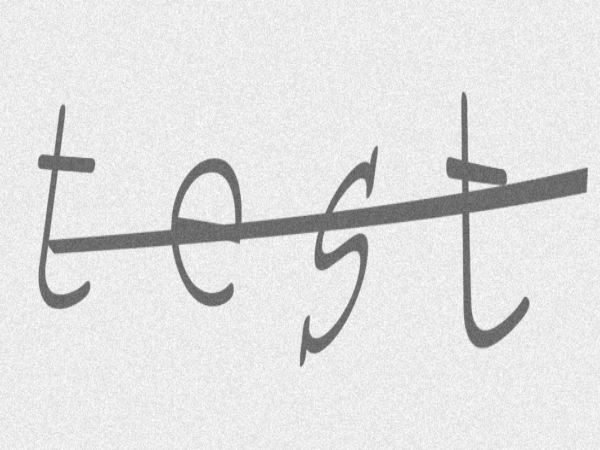} 
    \caption{Claptcha} 
    \label{fig_captchas:b} 
  \end{subfigure} 
  \begin{subfigure}[b]{0.2\linewidth}
    \centering
    \includegraphics[width=.7\linewidth]{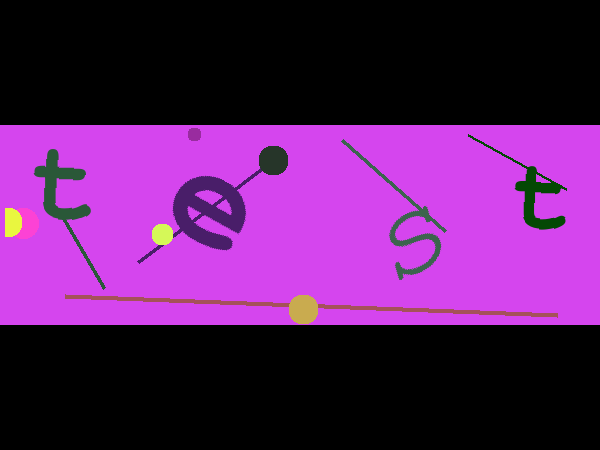} 
    \caption{Multicolor} 
    \label{fig_captchas:c} 
  \end{subfigure}
  \begin{subfigure}[b]{0.2\linewidth}
    \centering
    \includegraphics[width=.7\linewidth]{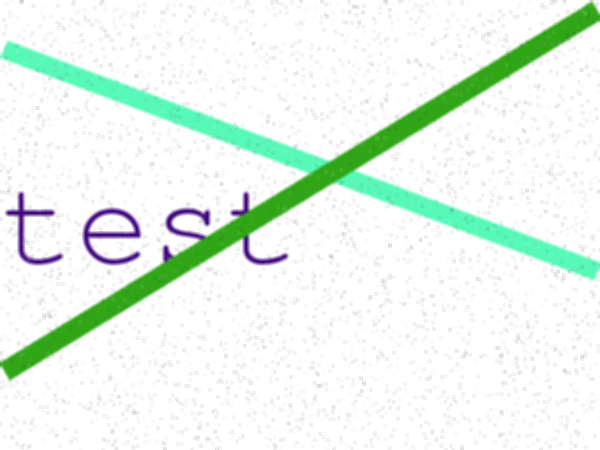} 
    \caption{Homemade} 
    \label{fig_captchas:d} 
  \end{subfigure} 
  \caption{Captchas' styles used in the experiments.}
  \label{fig_captchas} 
\end{figure*}

In this work, we are interested in understanding if ACM are vulnerable to textual captchas, particularly if different styles of textual captchas affect such target systems in different ways. We thus generate four variants of custom textual captchas. Each style differs in the type and number of transformations applied to the textual captcha. 
The four classes show different readability difficulties; the more transformations we apply, the more complex the image readability. 
We now describe the four adopted styles.
\begin{enumerate}[noitemsep,topsep=0pt,parsep=0pt,partopsep=0pt]
    \item \textit{Clean}. These are normal white images containing text. No further transformations are applied. Font: FreeMono.
    \item \textit{Claptcha}. Python captcha generator available on GitHub\footnote{\url{github.com/kuszaj/claptcha}}. Complex transformations are applied to the text. Font: FreeMono.
    \item \textit{Multicolor}. Python captcha generator available on GitHub\footnote{\url{github.com/J-Rios/multicolorcaptcha}}. We modified the library to use arbitrary text of arbitrary length. Complex transformations are applied to the text. Font: Free family fonts.
    \item \textit{Homemade}. Our captcha generator, it aims to be more readable than \textit{Claptcha} and \textit{Multicolor}. Simple transformations are applied to the text. Font: FreeMono.
\end{enumerate}
Figure~\ref{fig_captchas} shows examples of attacks, one per class. 
Table~\ref{tab:tranformations} summarizes the transformations applied in the four sample classes; we can notice that different styles adopt different transformations. For example, \textit{Clean} only uses only domain transfer, while \textit{Claptcha} and \textit{Multicolor} a high number of transformations.
We produce \textit{Clean} samples to verify if current deployed ACM deal with textual captchas, while we produced \textit{Claptcha}, \textit{Multicolor}, and \textit{Homemade} to verify if attackers can affect ACM' OCR.
About the domain transfer transformation, this operation is easy to implement, from graphic software (e.g., Paint, Photoshop) to standard programming libraries (e.g., matplotlib).
Similarly, all transformations described in Table~\ref{tab:tranformations} are easily applicable with the tools mentioned above. 
We further remark that there exist several online tools aiming to generate customisable textual captchas.
The aftermath is that even attackers with low computer skills can produce customisable undetectable textual captchas.
\par
The final dataset contains 2008 images. 
We do not make the dataset publicly available since it might be used for attacks in the real world. 
However, we make it available upon requests for researchers to facilitate future investigations in this field.

\begin{table}[!htb]
    \centering
    \scriptsize
    \caption{List of transformation functions for textual captchas variants.}
    \label{tab:tranformations}
    \setlength{\tabcolsep}{2pt} 
    \begin{tabular}{cccccc}
        \toprule 
         $T\#$ &\textbf{\textit{Transformation}}&\textbf{\textit{Clean}} &\textbf{\textit{Claptcha}}&\textbf{\textit{Multicolor}}&\textbf{\textit{Homemade}}\\
         \midrule
        $T_1$ & Domain transfer & \cmark & \cmark & \cmark & \cmark\\
        $T_3$ & Rotation & & \cmark & \cmark & \\
        $T_4$ & Distortion & & \cmark &  & \\
        $T_5$ & Waving & & \cmark &  & \\
        $T_6$ & Solid background & & & \cmark &\\
        $T_7$ & Noisy background & & \cmark &  & \cmark\\
        $T_8$ & Different fonts \& sizes & & \cmark & \cmark & \\
        $T_9$ & Different colors & &  & \cmark & \cmark \\
        $T_{10}$ & Occluding symbols & & \cmark & \cmark & \cmark \\\bottomrule
    \end{tabular}
\end{table}

\subsection{Captchas Readability}
\label{sec:user-study}
Ideally, if CAPA samples are posted on the web, they should be easy to read for humans, otherwise the whole attack would lose its purpose. 
Although posting unreadable content would surely evade any ACM, our samples need to have a good balance between low OCR-readability and high human-readability. 
While we evaluate efficacy of OCR in Section~\ref{sec.attack}, we assess our captchas human-readability through an user study.
We did not use any harmful word at this stage to not hurt anyone sensibility.

\paragraph{Methodology} 
From a list of of English verbs\footnote{github.com/aaronbassett/Pass-phrase}, we randomly selected 600 words, generating the corresponding custom textual captchas, 200 for each captcha class (\textit{Clapthca}, \textit{Multicolor}, \textit{Homemade}). 
Then, we asked 50 participants (27 females, 23 males, age mean 28.6, std 6.1) to annotated them. Each candidate annotated 50 samples, providing the text they could read along a difficulty score, from 1 (very easy) to 5 (very difficult), to express how much the participant was sure about the answer, and how immediate the captcha was to solve. The confidence score is crucial to understand if people are likely to read captchas while scrolling social networks feed, or would ignore them because considered difficult.
Each sample was processed by five participants. During the task, no time restrictions were given. 

\paragraph{Metrics}
Participants are evaluated with two metrics: accuracy and \textit{Character Error Rate} (CER). In particular, the accuracy evaluates the percentage of samples that were correctly annotated. CER, which is a popular OCR evaluation metric~\cite{handwrittenSurvey}, measure the character distance between the annotation and the ground truth (the lower, the closer are the two words).
The CER score is computed with Fastwer python library\footnote{github.com/kahne/fastwer}.

\begin{figure}[!htbp]
    \centering
    \includegraphics[width=.6\linewidth]{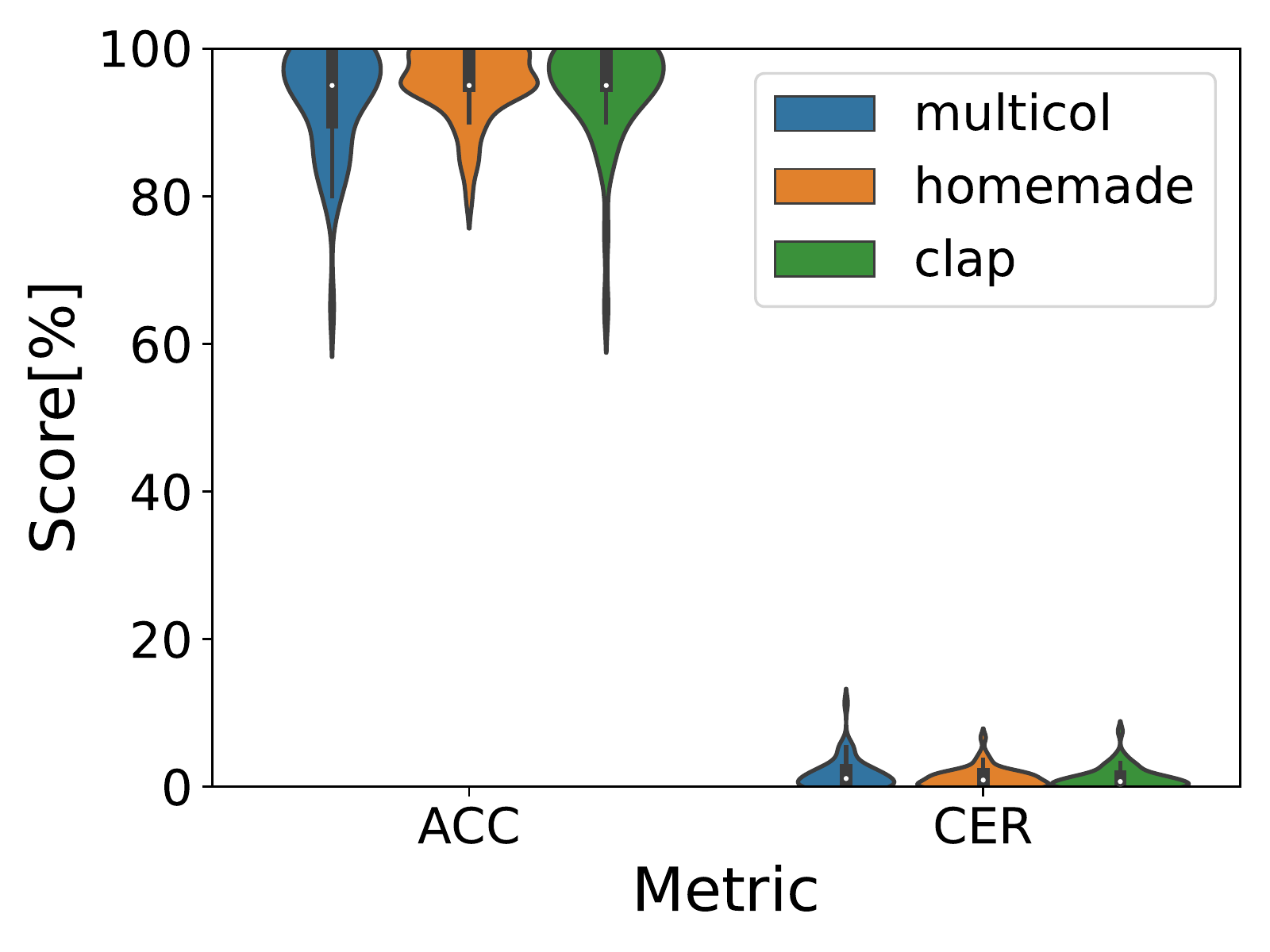}
    \caption{User-study performance distribution. We report accuracy (the higher, the better), and CER (the lower, the better).}
    \label{fig:readability}
\end{figure}

\paragraph{Results}
As shown in Figure~\ref{fig:readability}, we confirm the high readability of our samples.
On average, humans obtained 94.53\% and 1.31\% of accuracy and CER, respectively. Overall, the task was trivial, with low difficulty scores reported (\textit{Claptcha} = 1.3, \textit{Multicolor} = 1.5, \textit{Homemade} = 1.3).
Moreover, we counted the number of samples that have been always succesfully (or unsuccesfully) annotated by participants, producing and agreement score. Most samples are always correctly annotated (82.83\%), while only 0.5\% are always wrongly annotated.
We thus expect a comparable high readability on CAPA dataset as well.

\section{Attack Results}\label{sec.attack}
In this section, we present the results of our attacks in real-life scenarios.
We first discuss the attacking scenarios we consider (Section~\ref{sub.att_over}), followed by a presentation of the results of our attack against already deployed ACM (Section~\ref{sub.ar_c1}) and against ACM following the schema shown in Figure~\ref{fig:cross_domain_moderator} (Section~\ref{sub.ar_c2}).

\subsection{Overview}\label{sub.att_over}
Based on the discussions of ACM deployment done in Section~\ref{sub.am_as}, we aim to verify the following:
\begin{enumerate}[noitemsep,topsep=0pt,parsep=0pt,partopsep=0pt]
    \item Do current ACM consider cross-domain samples (e.g., text inside images) in their evaluations? We answer this question in Section~\ref{sub.ar_c1} by attacking image moderators with \textit{Clean} samples. We recall that these samples do not contain any transformation and, thus, OCR should successfully extract their text.
    \item If ACM consider cross-domain content, are they vulnerable to offensive textual captchas? We answer this question in Section~\ref{sub.ar_c2} by analyzing ACM responses on \textit{Clapcha}, \textit{Multicolor}, and \textit{Homemade} samples. 
\end{enumerate}
Tests of the ACM of social networks (e.g., Facebook) are not possible because it would imply the spread of inappropriate and harmful content. 
Furthermore, we cannot test the attacks to current state-of-the-art solutions (e.g., hateful memes detection) because, to the best of our knowledge, they all require that the text is successfully extracted through OCR~\cite{Kiela2020TheHM,Gomez_2020_WACV,velioglu2020detecting}.
Moreover, the hateful images presented in our dataset (see Section~\ref{sec.dataset}) contain only hateful text, while the rest of the background is not harmful. 
Thus, we opted to test already deployed ACM APIs. 


\subsection{Image Moderators}\label{sub.ar_c1}
This first experiment aims to verify that automatic content moderators do consider textual information in images during their prediction. 
To do so, we analyze the scores of only \textit{Clean} class samples. 
An example of \textit{Clean} image is shown in Figure~\ref{fig_captchas:a}.
%
We test the following ACM deployed by top IT companies.
\begin{itemize}[noitemsep,topsep=0pt,parsep=0pt,partopsep=0pt]
    \item \textit{Amazon Content Moderation}\footnote{\url{https://docs.aws.amazon.com/rekognition/latest/dg/moderation.html}}. The tool aims to classify inappropriate images among different classes, i.e., explicit nudity, suggestive, violent, visually disturbing, rude gestures, drugs, tobacco, alcohol, gambling, and hate symbols.
    \item \textit{Google Safe Search Detection}\footnote{\url{cloud.google.com/vision/docs/detecting-safe-search}}. The tool returns the likelihood of content containing spoof, medical, violent, or racy content. The likelihood is defined with the following classes: unknown, very unlikely, unlikely, possible, likely, and very likely. We consider content malicious if it is classified as possible, likely, or very likely.
    \item \textit{Microsoft Content Moderator}. The API identifies if the given image is appropriate for an adult audience (e.g., sexually explicit) or racist.
\end{itemize}
\par
%
%
%
%
%
%
For each analyzed ACM, we consider a post malicious if it is linked maliciously to at least one of the malicious classes. 
We measure the attack performance with the \textit{attack success rate} (ASR), defined as the ratio of unsafe content undetected divided by the total number of tests.
We find that all of the services cannot detect offensive text in images (without obfuscation). \textit{Clean} images reached a success rate of 1 for Amazon and Microsoft, and 0.96 for Google. 
The 4\% images labeled as inappropriate by Google were identified as \textit{spoofed}.
\par
This result suggests that analyzed CV-based ACM do not consider the case of images containing text. 
We highlight the gravity of such a finding: if an online platform adopts current ACM solutions, attackers could bypass their automatic monitoring systems by \textit{just} putting plain text inside images. To overcome this issue, online platforms should manually design defense mechanisms that follow, for example, the schema shown in Figure~\ref{fig:cross_domain_moderator}.
We believe that the ACM developers should address this issue instead since leaving uncovered our proposed scenario (text inside images) weakens their systems' reliability. 

%
%
%
%
%
\subsection{Cross-domain Moderators}\label{sub.ar_c2}
From the outcomes presented in Section~\ref{sub.ar_c1}, it seems that, by default, CV-based ACM do not consider textual info in images for their evaluation. 
To verify the power of textual captchas, we thus implement an ACM following the concepts introduced in Section~\ref{sub.am_as}.
In particular, we defined a pipeline that, given an image, extracts the text using an OCR, and then a textual ACM processes it. 
In this experiment, we vary the OCR technology while using Microsoft Content Moderator to spot potential harmful extracted sentences. 
We analyze OCR provided by Amazon\footnote{\url{aws.amazon.com/it/textract}}, Google\footnote{\url{cloud.google.com/vision/docs/ocr}}, Microsoft\footnote{\url{docs.microsoft.com/en-us/rest/api/cognitiveservices/contentmoderator/imagemoderation/ocrfileinput}}, and the popular free python library Tesseract\footnote{\url{pypi.org/project/pytesseract}}.
\par
We evaluate textual captchas with two metrics: the \textit{attack success rate} (ASR) as defined in Section~\ref{sub.ar_c1}, and the average \textit{normalized Levenshtein distance} (NLD):
\begin{equation}\label{eq.nld}
    NLD(x, x') = \frac{\mathcal{L}(x, x')}{max(|x|, |x'|)},
\end{equation}
where $x$ represents the true string in the image, $x'$ the OCR output, $\mathcal{L}$ the Levenshtein distance, and $|x|$ the number of characters in $x$. 
The Levenshtein distance measures the number of single-characters edits (e.g., addition, modification, deletion) required to make $x = x'$; it is defined between 0, when $x = x'$, and the maximum length between the two strings when they completely differ. 
The NLD measure defined in Equation~\ref{eq.nld} is thus defined in $[0, 1]$.
With the ASR we aim to understand the evasion power of our proposed attack, while with the NLD we aim to understand the number of mistakes that OCR do.
\par
\begin{figure}[!ht]
    \centering
    \includegraphics[width=0.9\linewidth]{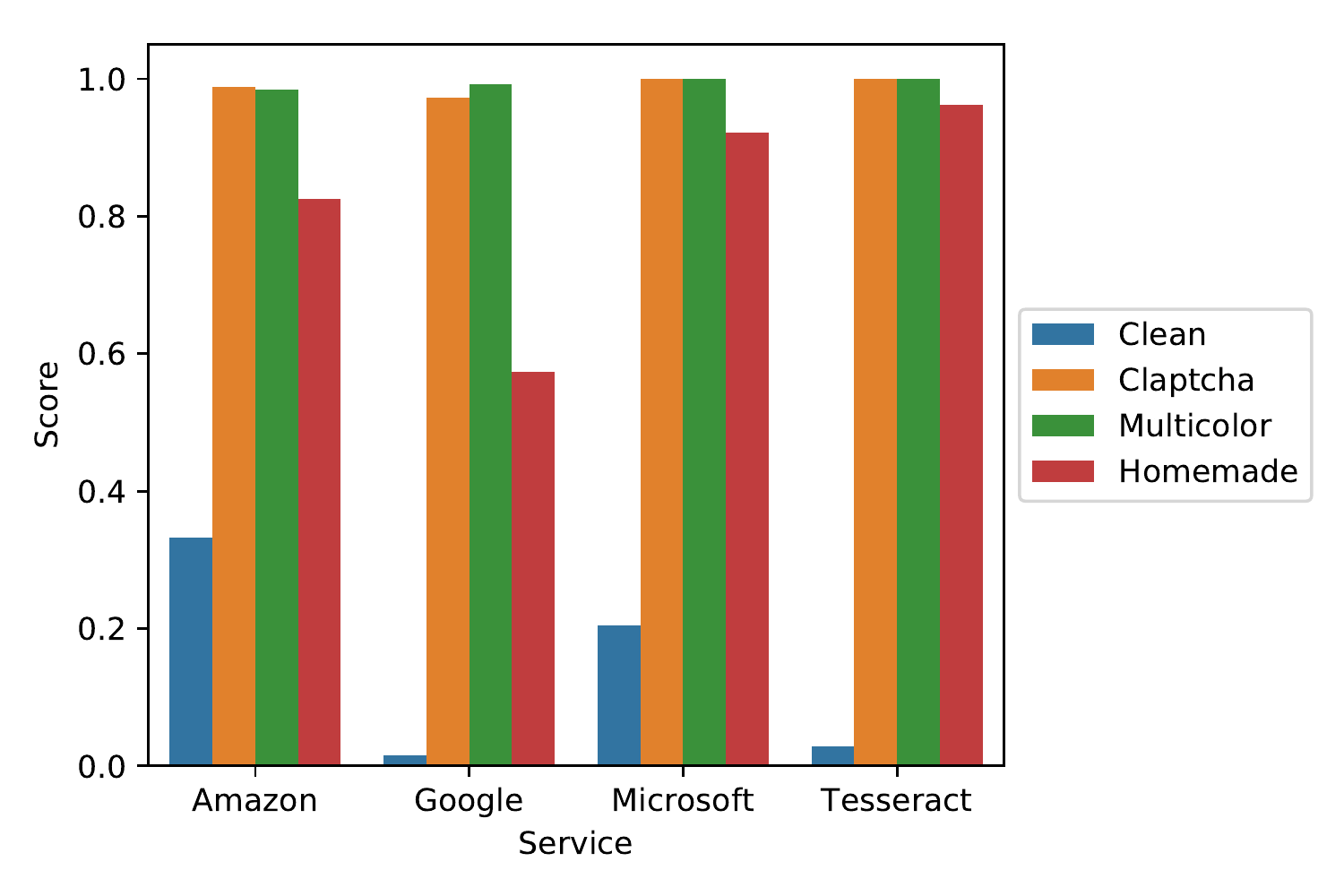}
    \caption{Cross-domain evaluation. We report the Attack Success Rate (ASR) (the higher, the better).}
    \label{fig:ASR}
\end{figure}

Figure~\ref{fig:ASR} shows the attack performance among the four services. 
We can first notice that the ASR rate on \textit{Clean} images is very low, meaning that OCR correctly extract the input text. 
We recall that \textit{Clean} samples do not have any visual transformation (e.g., rotation, complex background), and thus we expect that OCR work properly in such a case. 
This result suggests that ACM following the schema proposed in Section~\ref{sub.am_as} are resistant to those attacks that only apply the domain-transfer technique $T_1$.
Moreover, such a schema present a valid solution easily adoptable by commercial ACM. Indeed, the results on \textit{Clean} images are much higher compared to the one presented in Section~\ref{sub.ar_c1}.
\par
On the opposite, the ASR is close to 1.0 for both \textit{Claptcha} and \textit{Multicolor} variants, meaning that offensive textual captchas successfully evaded the ACM in all samples. 
Our captcha implementation \textit{Homemade} has an average ASR of 0.8, probably due to the less number of transformation applied compared to \textit{Claptcha} and \textit{Multicolor} (see Table~\ref{tab:tranformations}). 
Similar trends can be found with the NLD measure. We report such results in Appendix~\ref{app:attack}

\par
The results presented in this section suggest that ACM using the schema proposed in Figure~\ref{fig:cross_domain_moderator} could be vulnerable to textual captchas with few transformations (e.g., \textit{Homemade} class). Moreover, the more transformations, the higher the attack success rate, reaching the perfect evasion rate for \textit{Claptcha} and \textit{Multicolor}.

\section{Defense: Challenges \& Possible Directions}\label{sec.defense}
%
%

In this section, we present a countermeasure to CAPA. In particular, in Section~\ref{sub.def_overview} we discuss possible defense directions, followed by two implementations: supervised (Section~\ref{sub.supervised}) and unsupervised (Section~\ref{sub.unsupervised}) learning.

\subsection{Overview}\label{sub.def_overview}
\begin{figure*}[t!]
    \centering
    \includegraphics[width = 0.6\linewidth]{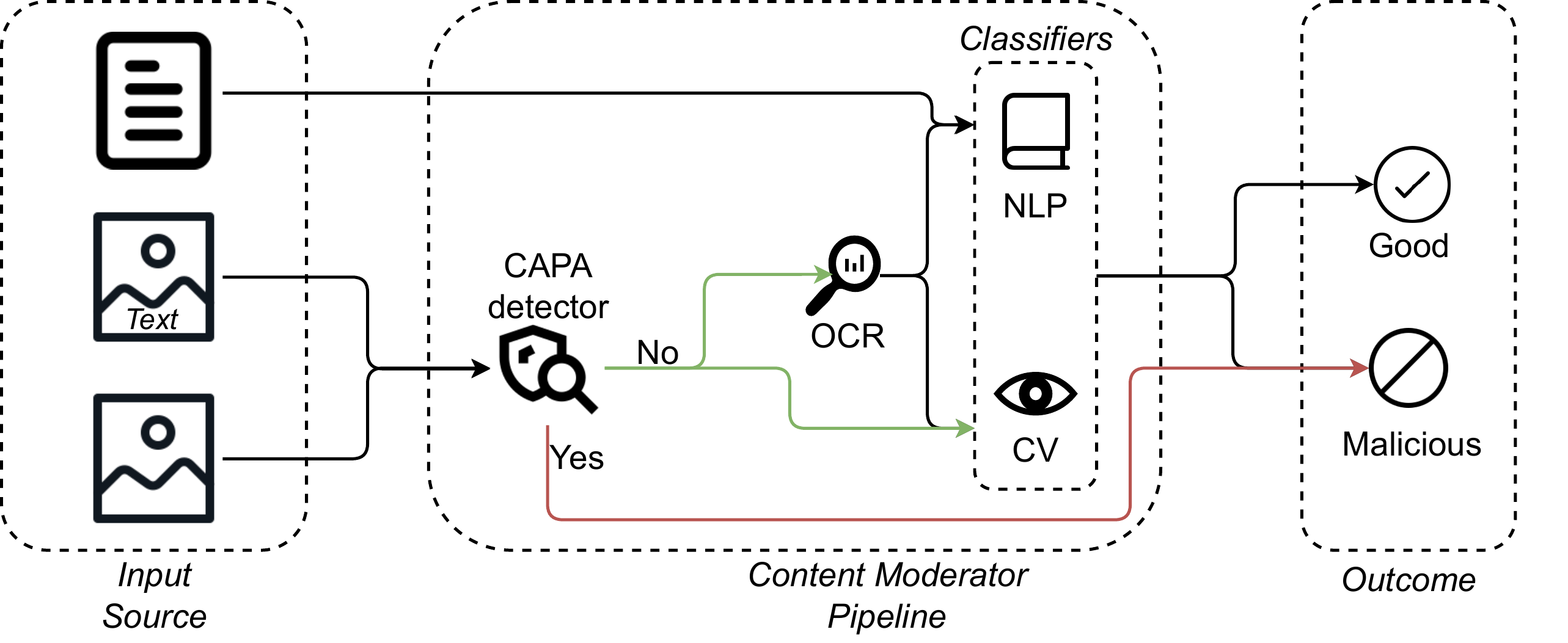}
    \caption{Overview of a content moderator in the text and image domains with our proposed countermeasure.}
    \label{fig:acm_defense}
\end{figure*}
In the previous section, we demonstrate how textual captchas can successfully evade ACM monitoring.
Since the generation process of customizable textual captchas is quite naive, this attack could be massively adopted by many users aiming to spread online messages without being censored. 
Indeed, as shown in Section~\ref{sec:taxonomy}, social network users are already adopting obfuscation techinques similar to CAPA samples.
Note that in the previous section, we tested the hate speech evasion task only for simplicity and to demonstrate the proposed attack capabilities. 
Nevertheless, the attack surface is not limited to hate speech evasion only, but can be extended to any text that an online platform can potentially ban, or more in general, analyze (e.g., opinion mining).
\par
It is thus necessary to discuss potential defenses and mitigation of our proposed attack CAPA. 
As we previously introduced in Section~\ref{sub.rw_cap}, captchas are generally a defensive mechanism. 
So far, the research community has primarily focused on the definition of new captchas or captchas breakers from an attacker's perspective.
The aftermath is that adopting textual captchas as an attack vector creates a novel and uncovered area of cyber security: the captchas identification. 
Indeed, captcha breakers start from the hypothesis to know a priori if an image is a captcha.
Thus, there is a need to tell if an image contains a textual captcha, or more in general, any obfuscation we discussed in our taxonomy (Section~\ref{sec:taxonomy}).
A CAPA detection mechanism should be integrated inside ACM. Whereas an image is recognized as a CAPA sample, the image should be immediately banned, or a human operator should check its content (i.e., an evasion attack is occurring). 
Figure~\ref{fig:acm_defense} shows the integration of the defense mechanism to the ACM schema introduced in Section~\ref{sub.am_as}. 

%
%
%
{\subsection{A Supervised Approach: Classification}\label{sub.supervised}}

\paragraph{Overview}
In Section~\ref{sub.def_overview} we motivated the need of a countermeasures to our proposed attack, and we identified a possible solution: the textual captcha identification.
We can model such a task as a binary classification problem, where the two classes are \textit{captcha} and \textit{non-captcha}. 
\paragraph{Dataset}
We now describe the datasets we used to deploy our defense, keeping in mind the following reasons:

\begin{enumerate}[noitemsep,topsep=0pt,parsep=0pt,partopsep=0pt]
    \item \textit{The target are OSN}. We must remember that, generally, ACM are deployed on OSN (e.g., Facebook, Twitter, Flickr). It is thus fundamental that the \textit{non-captcha} class captures representative data of the target OSN. 
    \item \textit{Unbalance dataset}. Intuitively, we might expect that the majority of the posts in an OSN are not CAPA samples. Thus, we expect the dataset to be unbalanced and that the \textit{non-captcha} class contains the majority of the samples.  
\end{enumerate}
We built three datasets, strarting from three distinct OSN for the \textit{non-captcha} class: Pinterest, Twitter, Yahoo-Flickr. We selected these datasets because images are a substantial portion of their daily content. 
For the \textit{captcha} class, we used the dataset the authors created in~\cite{ye2018yet}, made out of 11 different schemes, each with 700 samples, for a total of 7700 samples. We call this dataset C11.
In Appendix~\ref{app:styles}, more information about the captcha schemes contained in C11.
Table~\ref{tab:defense_dataset} summarizes the statistics of the four sources. 
We thus created the three datasets: Pinterest + C11, Twitter + C11, YFCC100M + C11.
Each dataset's version is split using 70\%, 10\%, and 20\% for the training, validation, and testing partitions. Due to computational limitations, we used just a random subset of YFCC100M.

\begin{table}[!htpb]
    \centering
    \footnotesize
    \caption{Statistics of the classes used for the evaluation of the defense.}
    \label{tab:defense_dataset}
    \begin{tabular}{ccc} \toprule
         \textbf{\textit{Origin}}  & \textit{\textbf{Class}} &\#\textbf{\textit{Samples} [$k$]} \\ \midrule
         C11~\cite{ye2018yet} & captchas & 7.7 \\
         Pinterest~\cite{pinterest} & non-captchas & 70\\
         Twitter~\cite{Vadicamo_2017_ICCVW} & non-captchas & 470\\ 
         YFCC100M~\cite{thomee2016yfcc100m} & non-captchas & 137\\
         \bottomrule
    \end{tabular}
\end{table}

\paragraph{Models}
In this work, we use two type of models: \textit{naive classifiers} and \textit{fine-tuned classifiers}. 
The naive classifier is defined as follows: Conv2D with kernel size = 5 and 6 output channels, followed by a second Conv2S with kernel size = 5 and out 16 output channels; the output is then flattened and forwarded to three linear layers (10K neurons, 1000, and 2 respectively). Each layer adopts the ReLU as the activation function; moreover, after both Conv2D we apply a MaxPool2D with kernel size = 2.  
For the fine-tuned models, we use three well known pre-trained models: Alexnet~\cite{krizhevsky2012imagenet}, Resnet18~\cite{resnet}, and VGG~\cite{simonyan2014very}.
The experiments are conducted in Pytorch. The fine-tuning strategy follows the official Pytorch tutorial\cite{pytorch-finetuning}. 
All models are trained using an SGD optimizer (learning rate = 0.001, momentum = 0.9), a cross-entropy loss, and an early stopping mechanism that stops the training if the validation loss is not optimized for five epochs. 
The models are trained for a maximum of 200 epochs. 

\paragraph{Results}
We evaluate our models using four standard metrics: F1-score macro, precision, recall, and ROC AUC. 
In general, all of the classifiers obtain strong classification results close to 100\% F1-score in all the scenarios (i.e., Pinterest + C11, Twitter + C11, YFCC100M + C11). 
This result implies that companies can easily recognize captchas schemes known at training time with appreciable performance. 
Detailed performance are available in Appendix~\ref{app:defense}.
\par
Despite these results, the definition of new captcha schemas is relatively easy by just varying the number and type of transformations (see Table~\ref{tab:tranformations}). Moreover, a specific type of transformation can be executed differently; for example, occluding symbols can vary (e.g., lines, segments).
We thus attempt to understand how well the trained models generalize to unknown captcha styles, i.e., captchas whose style is not present in the training set.
To do so, we test our models on the CAPA dataset (see Section~\ref{sec.dataset}), over the three classes \textit{Claptcha}, \textit{Multicolor}, and \textit{Homemade}. 
In 29 out of 36 cases (4 models * 3 classes * 3 datasets), the detection performance is equal to zero. 
This means that the defense does not generalize on new unknown captcha schemas. 
Only in a few cases, we have \textit{Claptcha} that are detected. 
A possible explanation is that \textit{Claptcha}' style is quite similar to some captchas styles presented in our training partition.
This confirm the need for the developing a more generalizable and reliable defense solution.

\subsection{An Unsupervised Approach: Outlier Detection}\label{sub.unsupervised}
\paragraph{Overview}
Supervised techniques guarantee high detection performance on known captcha schemes, while they poorly generalize on unseen styles. This is a major issue for a defense mechanism, especially when users can always add slight modifications to captcha styles, generating new ones.
This makes supervised approaches unreliable. 
As a result, we adopt an orthogonal perspective toward our problem. 
We can assume that CAPA is not exploited on the web platforms, and thus our samples might look different from regular posts that such platforms contain. 
Therefore, we adopt an outlier approach, where regular platforms posts are inliers and captcha outliers.




\paragraph{Dataset} 
We use the sources of the same datasets presented in Section~\ref{sub.supervised} but a different training and validation strategy.
In particular, the training set contains only samples belonging to the target OSN, while validation and test sets contain both benign and captcha samples.
For each OSN, we first take a random subset of 50K samples, and then we split it into training (70\%), validation (10\%), and testing set (20\%).
In our investigation, we are willing to understand \textit{how many captcha styles} we should know to build a robust defense. Thus, we vary the number of known captcha styles in the validation set based on the 11 classes available in the C11 dataset. We experiment with different $k$ known styles,  $k\in\{2, 4, 6, 8, 10\}$. For each scenario, we repeat the experiment with 5 different styles combinations.
The known captcha styles are then randomly split into validation and testing sets, with a 50\% of proportion. 
The unknown captcha styles will belong exclusively in the testing set. 

\begin{figure*}[!ht]
    \centering
    \includegraphics[width=.8\linewidth]{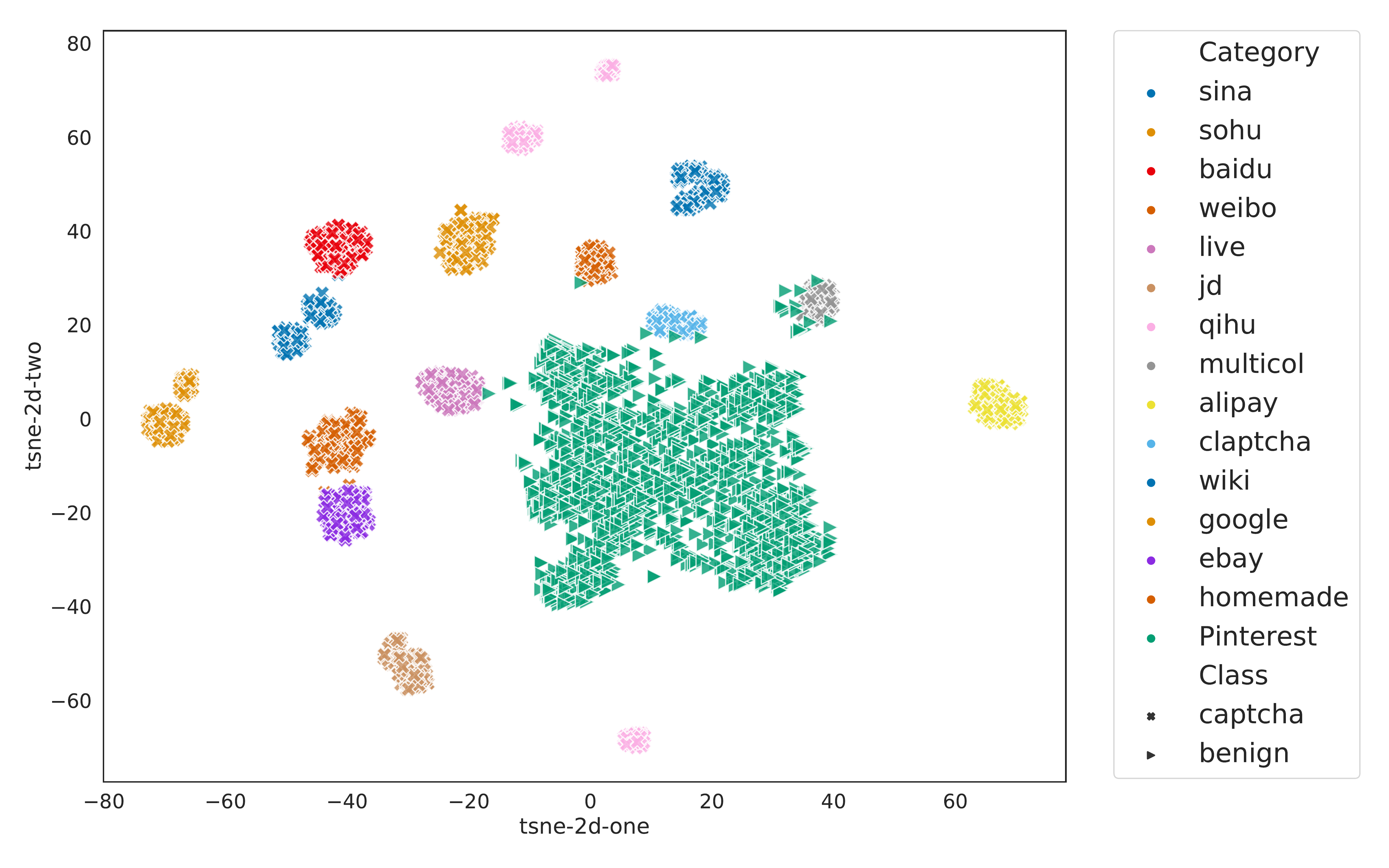}
    \caption{T-SNE 2D vislualization of 2000 samples benign (Pinterest) and 2000 captchas (C11 and CAPA).}
    \label{fig:pinterest}
\end{figure*}

\paragraph{Models}
All images are first converted into a 512-dimension embedding representation, using the pre-trained model ResNet-18~\cite{resnet}.
The first component of our defense is a dimensionality reduction module. We opted for the Principal Component Analysis (PCA), on which we vary the number of components: $[2, 8, 64, 128]$. 
We then tested the following algorithms: Isolation Forest (IF), Local Outlier Factor (LOF), ECOD~\cite{li2022ecod}, and One-Class SVM (OCSVM), using the implementation available in PyOD~\cite{zhao2019pyod}. For each model, we tune a common hyper-parameter, i.e., the contamination level $[0.1, 0.05, 0.01]$. Moreover, IF are tuned on the number of estimators $[16, 32, 64, 128]$, LOF on the number of neighbors $[2, 4, 8, 16]$, 
OCSVM on the kernel type $[rbf, sigmoid]$. 
All the models are tune with a grid-search strategy.

\begin{figure*}[!ht]
     \centering
     \begin{subfigure}[b]{0.3\textwidth}
         \centering
         \includegraphics[width=\textwidth]{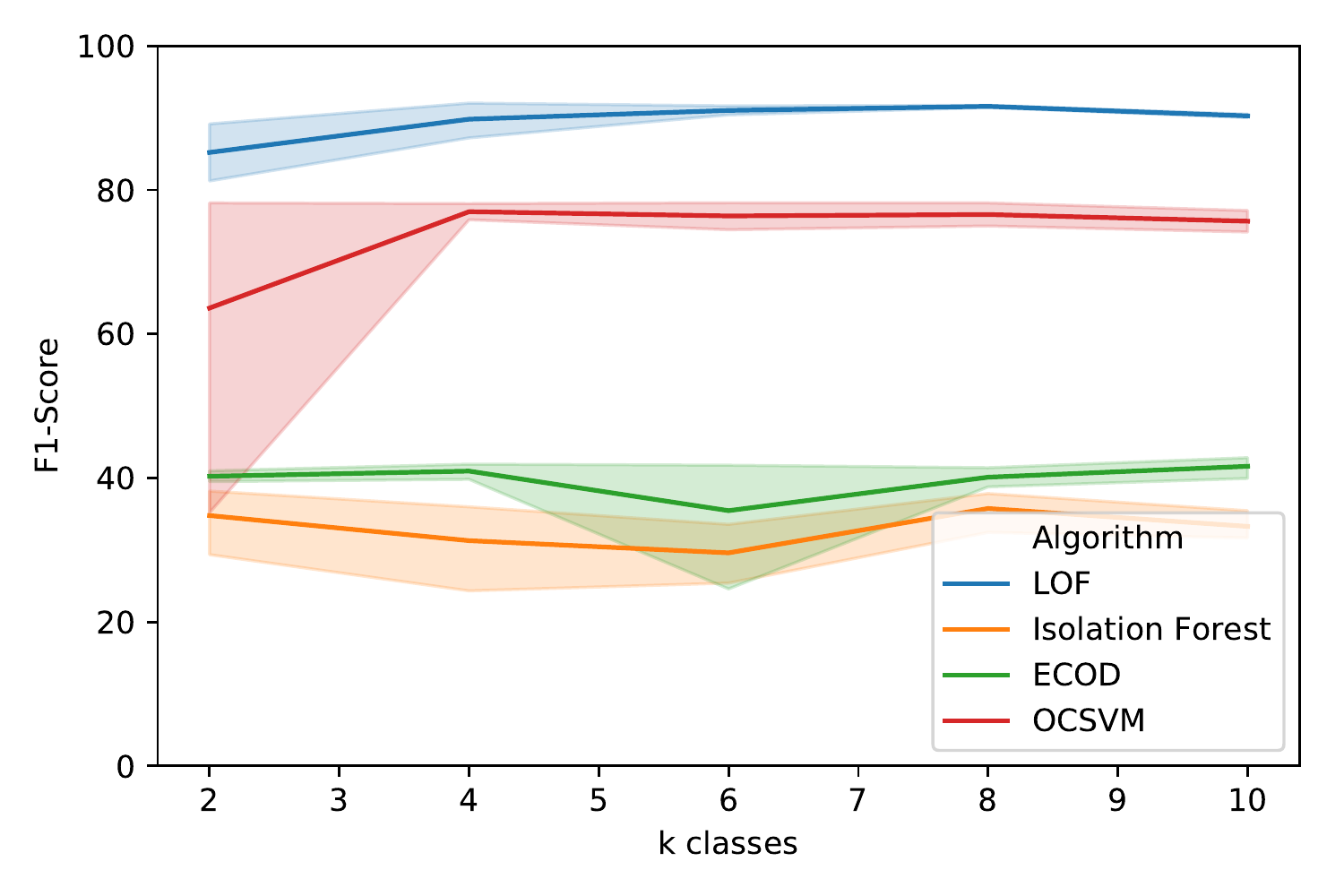}
         \caption{Pinterest.}
         \label{fig:od-test-Pinterest}
     \end{subfigure}
     \hfill
     \begin{subfigure}[b]{0.3\textwidth}
         \centering
         \includegraphics[width=\textwidth]{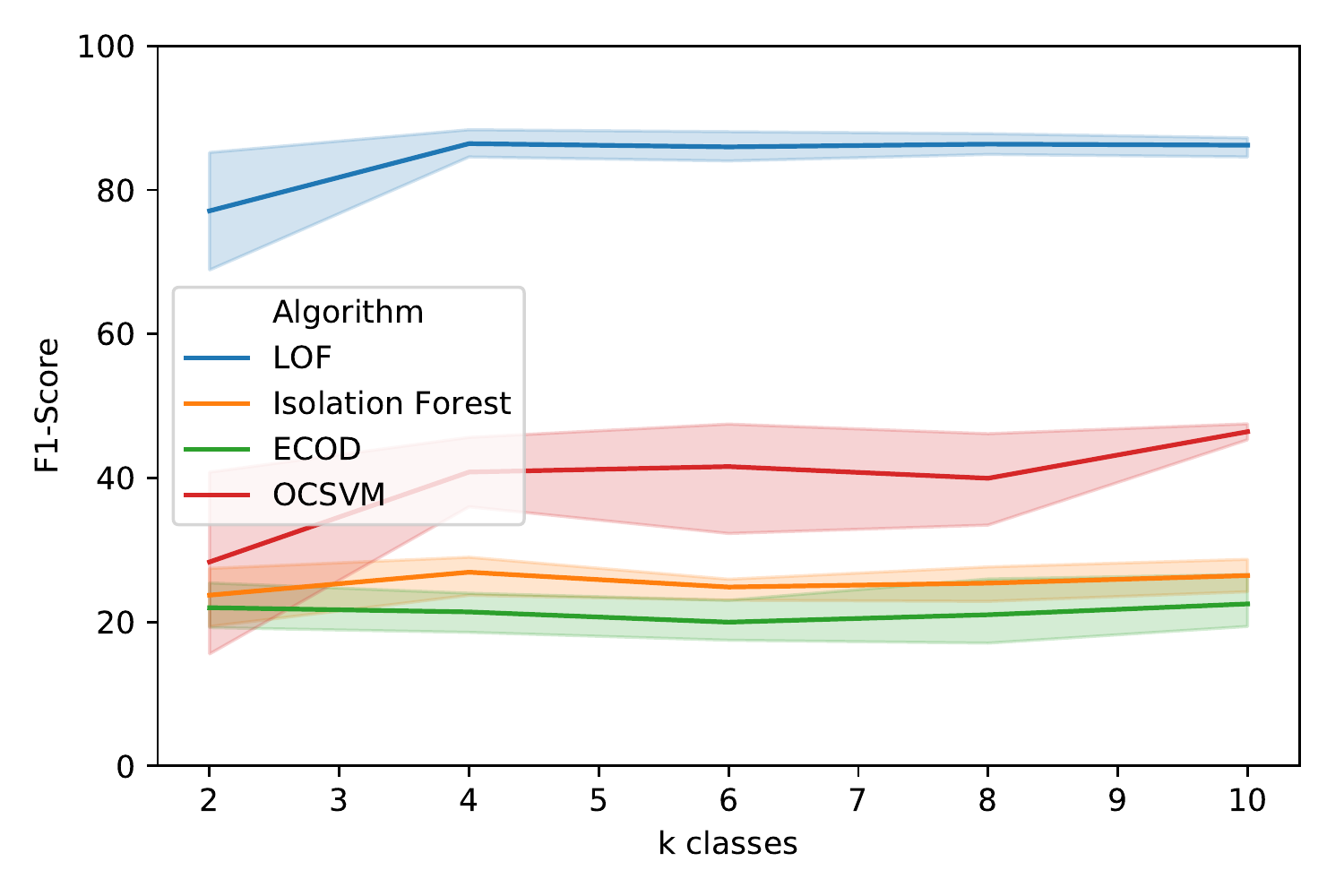}
         \caption{Twitter.}
         \label{fig:od-test-Twitter}
     \end{subfigure}
     \hfill
     \begin{subfigure}[b]{0.3\textwidth}
         \centering
         \includegraphics[width=\textwidth]{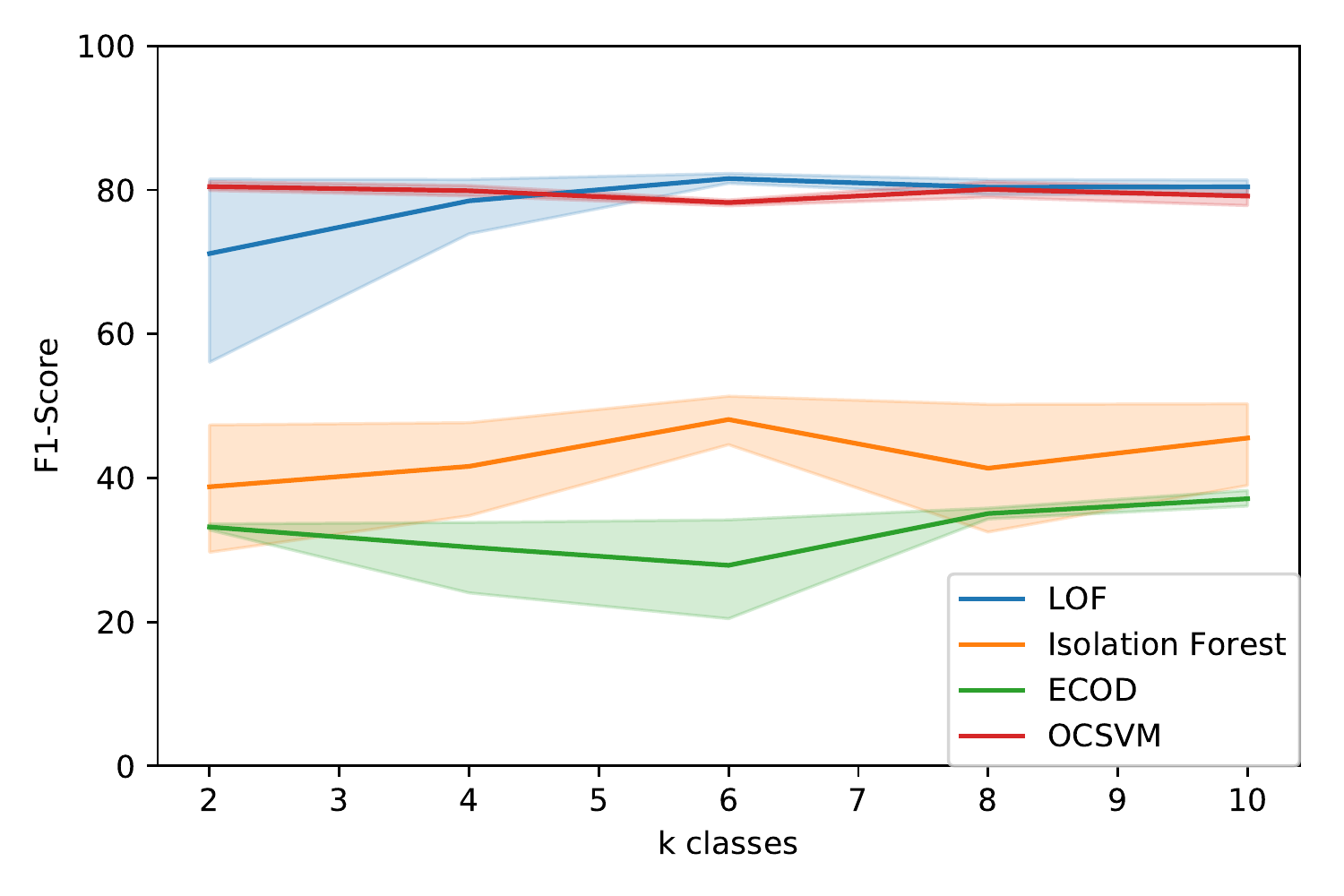}
         \caption{Yahoo-Flickr.}
         \label{fig:od-test-YF}
     \end{subfigure}
        \caption{F1-score of different Outlier Detection at the varying of the OSN.}
        \label{fig:od-test}
\end{figure*}

\paragraph{Results}
We first visually analyze our data, to better understand possible outcomes. 
Let consider the combination of Pinterest, C11, and CAPA datasets. We randomly sampled 2000 items each. From these samples, we first extracted the embedding, and obtained a two dimensional feature space with the combination of a PCA (from 512 to 50 features) and T-SNE (from 50 to 2 features).
Figure~\ref{fig:pinterest} shows how the samples are distributed. We can notice that captchas samples have distinct and unique pattern compared to Pinterest ones.  
However, each captcha styles defines an own and distinct cluster as well, explaining the poor generalization performance in classification tasks. 
We now instead move on the results of the three outlier detectors. Figure~\ref{fig:od-test} shows the F1-score at testing time at the varying of the number of known captcha styles used in the validation set. 
LOF outstands both Isolation Forest, ECOD, and OCSVM in the three OSN scenarios, reaching, on average a performance of 80\% F1-score. 
Moreover, we can notice that the amount of known styles has a limited impact, finding a performance stabilization starting from 4 styles. 
Furthermore, we identify consistent trends with both known and unknown captcha styles recognition. More details in Appendix~\ref{app:defense}.
The presented results suggests that LOF might be a suitable tool to use `into the wild'. 
The major limitation of our approach is the high number of false positive, which we estimate close to 9\% of benign posts, on average. 
By visually analyzing the false positive, we discover that, on Pinterest 50\% contains textual information, and the 30\% presents text over hard background. On Twitter, we found that 63\% of anomalies contain text, and 24\% contain text over hard background.  On Yahoo-Flicker, only the 5\% of anomalies instead contain text. 
In all three cases, only a few anomalies are linked to other obfuscation we identified in Section~\ref{sec:taxonomy}.

\section{Conclusions and Future Works}\label{sec.Conclusions}
Content moderators play an essential role in nowadays society for the moderation of inappropriate content spread and shared on online platforms. 
Dangerous content (e.g., hateful words, nudity images) can potentially reach a broad audience, hurting or harming sensitive people.    
Online platforms started adopting automatic tools based on deep learning solutions to deal with the massive content volume. 
\par
As part of this work, we first presented a taxonomy of new attacks perpetrated by OSN users in order to evade ACM. 
Inspired by these samples, we presented CAPA, an attack that leverages custom textual captchas to evade such detection tools.
Our results prove the high efficacy of our proposed attack. 
We demonstrated how easily an attacker could elude ACM detection by i) changing the domain from text to image and ii) applying captchas schemes. 
With the first, an attacker can evade those ACM not considering images containing text scenarios.
With the latter, an attacker can affect NLP-based tools' performance by exploiting OCR' weaknesses. 
While CAPA is easy to implement and does not require any information about the target model, identifying a countermeasure seems challenging due to generalizability issues. 
Toward this direction, we propose an effective defense based on outlier detection reaching 80\% F1-score. 
\par
Our work poses several challenges that might inspire future works. 
First of all, it is necessary to define a boundary between captchas and non-captchas. 
Even the other samples of our taxonomy can be considered as textual captchas, since their overall goal is to deceive OCR. 
Second, for the various categories of the proposed taxonomy, a proper dataset should be collected, to eventually train detectors or sanitizers to help ACM.
Last, it would be ideal to build a model that works against all the obfuscation variants described in taxonomy (e.g., emoji, leet speech). We believe that the computer-vision community can help in this direction by clarifying and proposing ad-hoc solutions for the problems we raised. 

\bibliographystyle{IEEEtran}
\bibliography{main}
\appendix
\subsection{Captcha Schemes}\label{app:styles}
During our experiment, we used the dataset from~\cite{ye2018yet} to represent the captcha class. Table~\ref{tab:captchas} shows examples of them along with the applied transformations.

\begin{table*}[!ht]
    \centering
        \footnotesize
        \def\arraystretch{1}
        \caption{Captcha schemes used in our experiment coming from~\cite{ye2018yet}.}
    \label{tab:captchas}
    \begin{tabular}{m{2.5cm} m{5cm} m{4.5cm}}
    \toprule
    \textbf{\textit{Scheme}} & \textbf{\textit{Example}} & \textbf{\textit{Trasformations}}\\
    \midrule
        Alipay &  \includegraphics[width = 0.5\linewidth]{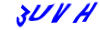} & Overlapping, rotation, distortion\\
        Baidu &  \includegraphics[width = 0.5\linewidth]{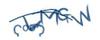} & Occluding lines, overlapping, rotation, distortion, waving, varied font size \& color\\
        eBay &  \includegraphics[width = 0.5\linewidth]{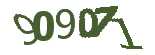} & Overlapping, distortion, rotation, waving\\
        Google &  \includegraphics[width = 0.5\linewidth]{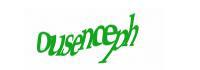} & Overlapping, rotation, distortion, waving, varied font sizes \& color\\
        JD &  \includegraphics[width = 0.5\linewidth]{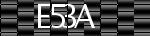} & Overlapping, rotation, distortion\\
        Microsoft &  \includegraphics[width = 0.5\linewidth]{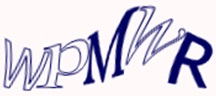} & Overlapping, solid background, rotation, waving, varied font syles \& sizes\\
        Qihu360 &  \includegraphics[width = 0.5\linewidth]{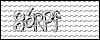} & Overlapping, rotation, distortion, varied font sizes\\
        Sina &  \includegraphics[width = 0.5\linewidth]{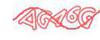} & Overlapping, rotation, distortion, waving\\
        Sohu &  \includegraphics[width = 0.5\linewidth]{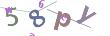} & Overlapping, complex background, occluding lines, rotation, varied font size \& color\\
        Weibo &  \includegraphics[width = 0.5\linewidth]{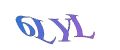} & Overlapping, occluding lines, rotation, distortion\\
        Wikipedia &  \includegraphics[width = 0.5\linewidth]{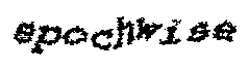} & Overlapping, rotation, distortion, waving\\
    \bottomrule
    \end{tabular}

\end{table*}

\subsection{Attack}\label{app:attack}
In Figure~\ref{app:attack}, the performance of CAPA among different targets.
\begin{figure*}[!ht]
     \centering
     \begin{subfigure}[b]{0.48\textwidth}
         \centering
         \includegraphics[width=0.9\textwidth]{Figures/ASR_.pdf}
         \caption{ASR}
         \label{fig:app-ASR}
     \end{subfigure}
     \hfill
     \begin{subfigure}[b]{0.48\textwidth}
         \centering
         \includegraphics[width=0.9\textwidth]{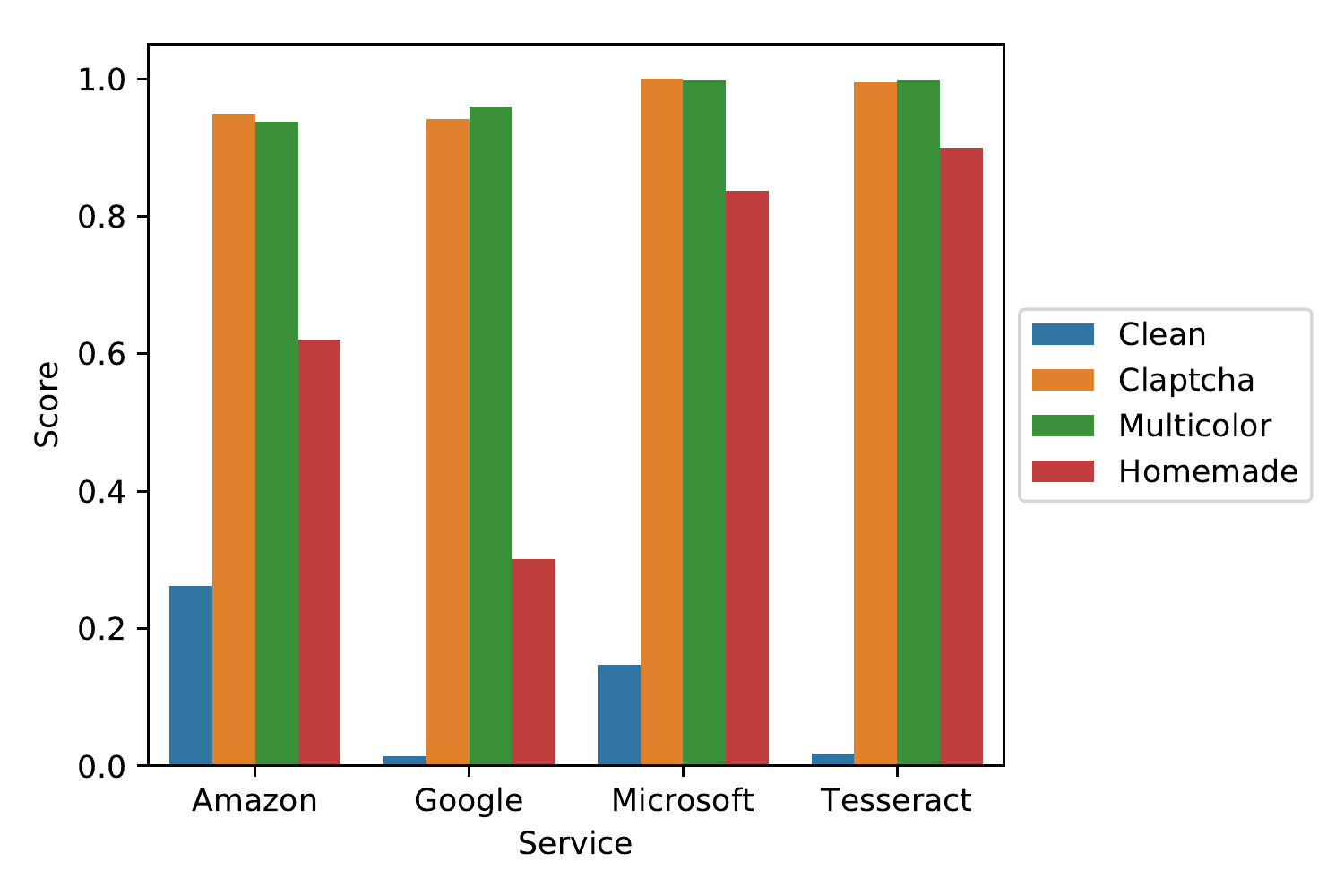}
         \caption{NLD}
         \label{fig:app-NLD}
     \end{subfigure}
        \caption{Cross-domain evaluation. On the left, the Attack Success Rate (ASR). On the right, the average Normalized Levenshtein Distance (NLD). For both measures, the higher, the more successful the attack.}
        \label{fig:app-attack2}
\end{figure*}

\subsection{Defense}\label{app:defense}
Table~\ref{tab.fidelity_novelty} summarizes the results of the classifiers when the captcha styles are known.
Table~\ref{tab.transferab} shows the generalization performance over the three CAPA classes: \textit{Claptcha}, \textit{Multicolor}, and \textit{Homemade}.
\begin{table*}[ht!]
\centering
\footnotesize
\def\arraystretch{1}
\caption{Detection results of the models in different OSNs. For each metric higher scores defines better classifiers.}
 \label{tab.fidelity_novelty}
  \begin{tabular}{lcccc| cccc| cccc}
    \toprule
        
        \multicolumn{1}{l}{\textbf{\textit{Dataset}}} &   \multicolumn{4}{c}{\textbf{\textit{Pinterest}}} &
        \multicolumn{4}{c}{\textbf{\textit{Twitter}}} &
        \multicolumn{4}{c}{\textbf{\textit{YFCC100M}}} 
        \\
             
      \cmidrule(lr){2-5} \cmidrule(lr){6-9}  \cmidrule(lr){10-13}
      
    \multicolumn{1}{l}{\textbf{\textit{Metrics}}}  
    & F1 & Prec. & Rec. & AUC
    & F1 & Prec. & Rec. & AUC
    & F1 & Prec. & Rec. & AUC
     \\
    \hline

    Naive &   99.8 & 99.3 & 99.9 & 99.9 &	99.5 &	99.2 &	99.2 &	99.5 & 99.9 &	99.6 &	100	& 99.9 
          \\ 
    
    Alexnet	& 99.9 &	99.8 &	1 &	99.9 & 	99.8 &	99.5 &	100 &	100 &	99.9 &	99.8	& 100 &	100 
    \\
    Resnet18 &	100 &	100 &	100 &	100 & 99.9 &	99.7 &	100 &	100 &	99.9 &	99.7 &	100	& 100	
    \\
    VGG &	99.9 &	99.9 &	100 &	100 & 99.9 &	99.6 &	100 &	100 &	99.9 &	99.8 &	100 &	99.9	
    \\

    \bottomrule
\end{tabular}
\end{table*}

\begin{table*}[!ht]
\centering
\footnotesize
\def\arraystretch{1}
\caption{Percentage of CAPA captchas detected by models trained on data coming from different OSNs.}
 \label{tab.transferab}
  \begin{tabular}{lccc| ccc| ccc}
    \toprule
        
        \multicolumn{1}{l}{\textbf{\textit{Dataset}}} &   \multicolumn{3}{c}{\textbf{\textit{Pinterest}}} &
        \multicolumn{3}{c}{\textbf{\textit{Twitter}}} &
        \multicolumn{3}{c}{\textbf{\textit{YFCC100M}}} 
        \\
             
      \cmidrule(lr){2-4} \cmidrule(lr){5-7}  \cmidrule(lr){8-10}
      
    \multicolumn{1}{l}{\textbf{\textit{Classes}}}  
    & Clap & Multicol & Homemade
    & Clap & Multicol & Homemade
    & Clap & Multicol & Homemade
    \\
    \hline
    Naive &   11.95 &	0 &	0 &		1.2 &	0 &	0 & 		51.79 &	0 &	0.2 \\ 
    
    Alexnet	& 11.16 &	0 &	0 &		1 &	0 &	0 &		52.19 &	0 &	0
    \\
    Resnet18 &	0 &	0 &	0 &		0 &	0 &	0 &		0 &	0 &	0
    \\
    VGG&	0 &	0 &	0 &		0 &	0 &	0 &		0 &	0 &	0    \\
    \bottomrule
\end{tabular}
\end{table*}

\begin{figure*}[!ht]
    \centering
    \includegraphics[width = .8 \linewidth]{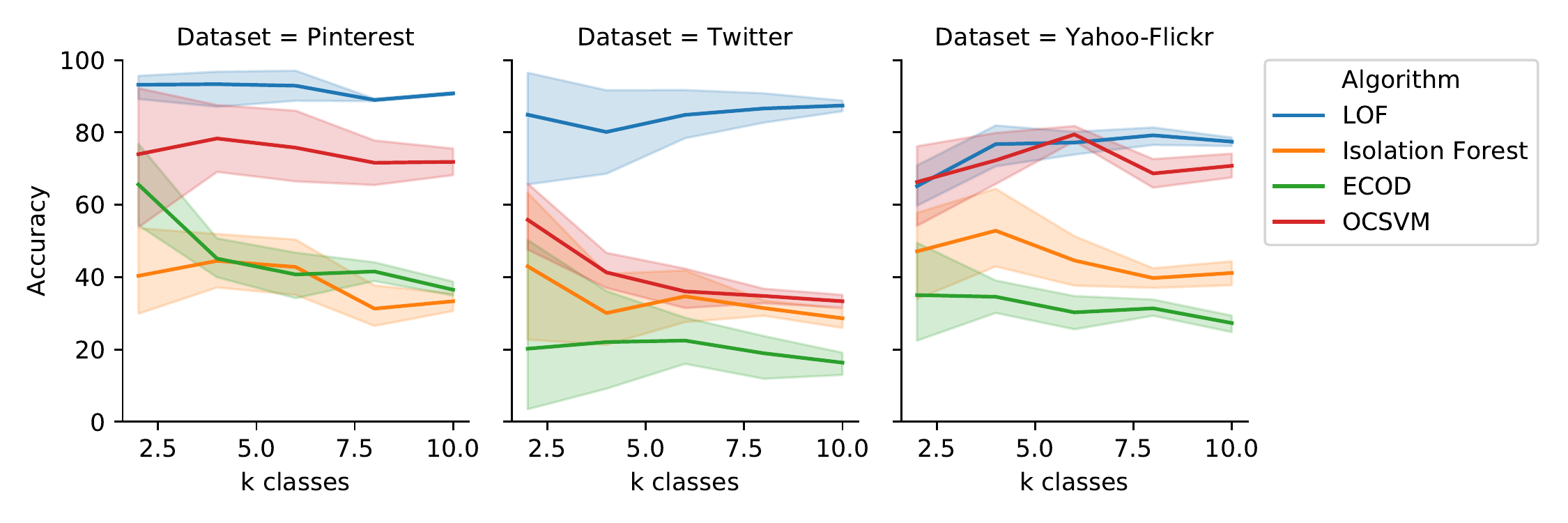}
    \caption{Accuracy of different Outlier Detection on known captcha styles at the varying of the OSN.}
    \label{fig:od-k}
\end{figure*}

\begin{figure*}[!ht]
    \centering
    \includegraphics[width = .8 \linewidth]{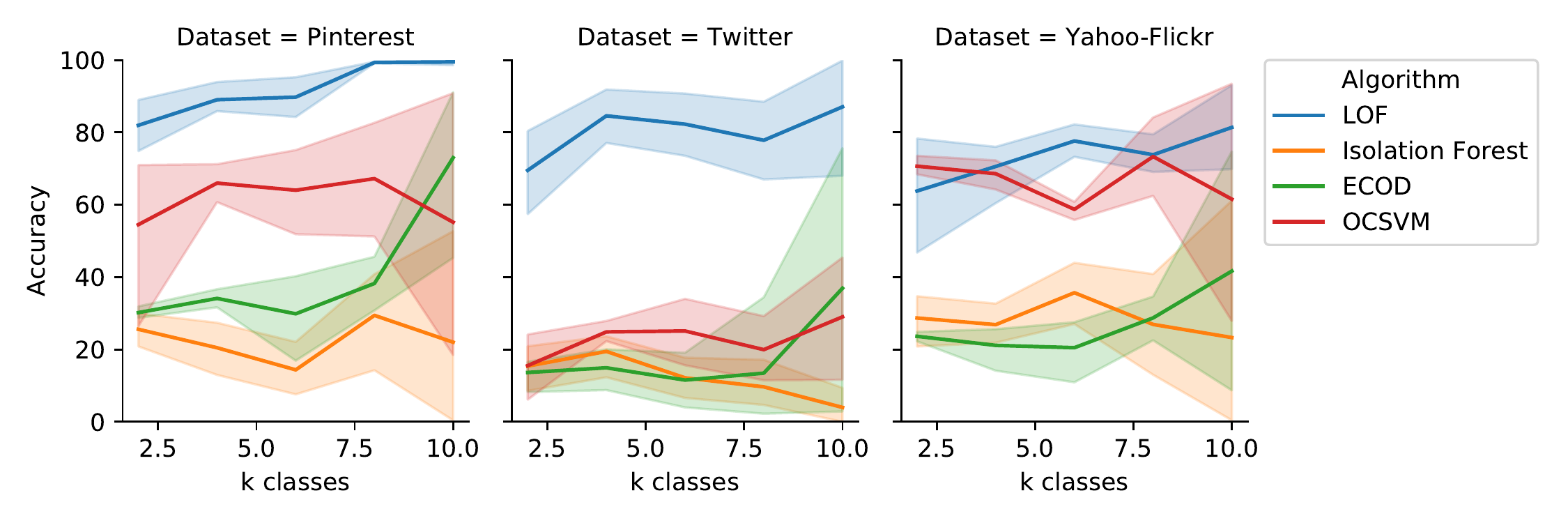}
    \caption{Accuracy of different Outlier Detection on unknown captcha styles at the varying of the OSN.}
    \label{fig:od-uk}
\end{figure*}

\end{document}